\documentclass[%
aip,
amsmath,amssymb,
reprint,%
]{revtex4-1}

\usepackage{graphicx}
\usepackage{dcolumn}
\usepackage{bm}

\usepackage[utf8]{inputenc}
\usepackage[T1]{fontenc}
\usepackage{mathptmx}
\usepackage{etoolbox}
\usepackage{upgreek}
\usepackage{xcolor}

\usepackage[english]{babel} 

\newcommand{\blue}[1]{\textcolor{black}{#1}}

\makeatletter
\def\@email#1#2{%
	\endgroup
	\patchcmd{\titleblock@produce}
	{\frontmatter@RRAPformat}
	{\frontmatter@RRAPformat{\produce@RRAP{*#1\href{mailto:#2}{#2}}}\frontmatter@RRAPformat}
	{}{}
}%
\makeatother
\begin{document}
	
	\preprint{AIP/123-QED}
	
	\title[All-optical compact setup for generation of collimated multi-MeV proton beams with a ”snail” target]{All-optical compact setup for generation of collimated multi-MeV proton beams with a ”snail” target}
	\author{N. Bukharskii}
	\altaffiliation[Also at ]{P.~N.~Lebedev Physical Institute RAS, Moscow, Russia.}
	\author{Ph. Korneev}%
	\altaffiliation[Also at ]{P.~N.~Lebedev Physical Institute RAS, Moscow, Russia.}
	\email{korneev@theor.mephi.ru}
	\affiliation{ 
		National Research Nuclear University MEPhI, Moscow, Russia
	}%


\begin{abstract}{The work considers an optical scheme for collimation of high-energy proton beams using $\sim 10^5$~T scale magnetic fields induced in a miniature "snail" target by petawatt or multi-petawatt laser irradiation in ps or fs regime. Such magnetic fields are known to be frozen into hot plasma and exist on at least a hundred of picoseconds time-scale, allowing their use for control of charged particle beams. The high values of the magnetic field along with the compact size perfectly match conditions for an all-in-one optical setup, where first, the laser beam accelerates protons, by, e.g. Target Normal Sheath Acceleration (TNSA) mechanism, and second, the closely positioned snail target is driven to guide the proton beam. An important issue is that the laser drivers for both proton acceleration schemes and the magnetic field generation in the considered targets may have the same properties, and even be parts of one splitted beam. Numerical simulations show that the considered setup can be used for efficient collimation of $\simeq 100$~MeV protons. The collimation effect weakly depends on the fine magnetic field structure and can be observed both for a simple magneto-dipole field profile and for a more complex coaxial-like profiles accounting for the intricate structure of electric currents in the interaction region. The obtained results are interesting for the development of intense laser-driven sources of charged particle beams with low divergence and high energy of accelerated particles.}
\end{abstract}




\maketitle

\section{Introduction}\label{sec1}

With the development of technologies for generation and amplification of ultra-short laser pulses, their intensities far exceeding $10^{18}$~W/cm$^2$ became attainable~\cite{Yoon_Optica_2021}. When such intense laser fields interact with matter, they can effectively accelerate electrons \cite{Mourou_RevModPhys_2006, Pukhov_PhysPlasmas_1999, Pukhov_Review_2003, Rosmej_PPCF_2020, Esarey_RevModPhys_2009} and ions~\cite{Maksimchuk_PRL_2000, Snavely_PRL_2000, Roth_2016}. In the latter case, the acceleration nature usually is the charge separation fields formed in the process of laser irradiation of a solid target. Laser-driven sources of high-energy charged particles can offer several advantages over other schemes based on conventional accelerators. They present a very compact setup where strong accelerating fields allow reaching very high acceleration gradients~\cite{Katsouleas_IEEE_1997}. At the same time, the number of laser-accelerated particles can also be quite high both for electron~\cite{Rosmej_PPCF_2020} and ion~\cite{Roth_2016} acceleration. Laser-accelerated electron beams carrying high charges and high electric currents can be used for intense gamma-ray and neutron production~\cite{Gunther_NatCommun_2022, Tavana_Frontiers_2023}, research in the field of nuclear photonics~\cite{Ma_MRE_2019, Nedorezov_2021}. In addition, directed beams of accelerated electrons can be used as a driver for ion acceleration~\cite{Roth_2016}. Laser-accelerated ions and, in particular, protons, can be used e.g. as a diagnostic tool for probing ultra-fast transient phenomena, like pulsed electromagnetic fields generated with powerful lasers~\cite{Willingale_PhysPlasmas_2010, Law_APL_2016, Gao_PhysPlasmas_2016, Palmer_PhysPlasmas_2019, Bradford_HPLSE_2020}, and to produce warm dense matter for equation-of-state studies~\cite{Patel_PhysRevLett_2003, Dyer_PhysRevLett_2008}. Fast protons are also of a great interest in the context of fast ignition approach to inertial confinement fusion~\cite{Temporal_PhysPlasmas_2002, Bychenkov_PlasmaPhysRep_2001}. Also, laser-accelerated protons offer a promising potential for medical applications, i.e. cancer treatment with the hadron therapy~\cite{Bulanov_PhysLettA_2002}. 

Despite their advantages, laser-driven sources of charged particle beams typically suffer from a poor beam quality and a large divergence in comparison to conventional accelerators, limiting their use for potential applications such as the radiation therapy~\cite{Fourkal_MedPhys_2002, Fourkal_MedPhys_2003}. It is, however, possible to reduce the beam angular spread by exploiting strong electromagnetic fields for beam collimation. For example, it has been shown~\cite{Bailly-Grandvaux_NatCommun_2018, Santos_PhysPlasmas_2018} that laser-induced magnetic fields of about $\left( 500-600 \right)$~T can be used to magnetize a solid target and achieve a guiding of relativistic electron beams generated by irradiation of a target surface with a high-energy picosecond laser pulse. In this case, interaction with the magnetic field in the process of propagation through the magnetized target decreased the electron beam divergence and caused a 5-fold increase of the energy-density flux at the rear edge of a $60$-$\upmu$m-thick target~\cite{Bailly-Grandvaux_NatCommun_2018, Santos_PhysPlasmas_2018}. Strong laser-driven electromagnetic fields have also been applied for ion collimation, see Refs.~\cite{Kar_NatCommun_2016, Ahmed_SciRep_2017, Liu_PhysPlasmas_2024}.
There, travelling electromagnetic pulses were excited in helical coils and the fields associated with those pulses were shown to be capable of a guided post-acceleration of $\sim \left( 1-10 \right)$~MeV TNSA-produced protons. 

A practical point for applications is the simplicity and robustness of a possible setup which produces a collimated proton beam. So called capacitor-coil targets \cite{Santos.etal_LaserdrivenPlatformGeneration_NJP-2015, Santos_PhysPlasmas_2018,Law-apl16,Daido1986,Courtois-jap05} are well studied in ns - kJ regime, where they are able to create kilotesla-scale magnetic fields for several ns. However, in a ps regime, their efficiency seems to be much lower, see, e.g. \cite{Ehret.etal_GuidedElectromagneticDischarge_PoP-2023}. This means that in order to build an optically controlled compact source of collimated protons with use of capacitor-coil targets, probably two kinds of laser channels are required: the one for a high-power short (fs or ps) beam, and the other for a high-energy long (ns or hundreds of ps) beam. There are possible solutions aiming to simplify a possible setup, e.g. the travelling discharge, discussed in e.g. \cite{Ahmed_SciRep_2017}, may use a short powerful beam, even the same which creates the TNSA protons.

As an alternative to the setups with electromagnetic pulses in a helical coil, the beam collimation may be achieved with the use of strong quasi-stationary magnetic fields excited in miniature "snail" targets in the picosecond \cite{Ehret_PhysRevE_2022} or even in a femtosecond~\cite{Bukharskii_BullLebedevPhysInst_2023} regime. This approach requires two laser beams but allows for a possibility of an additional adjustment freedom. With a relatively modest $50$~J of invested laser energy in the form of sub-petawatt $\simeq 0.5$~ps laser pulse, the magnetic field strength in such a target can reach kilotesla level~\cite{Ehret_PhysRevE_2022}, while its life-time exceeds $\simeq 100$~ps. With more powerful laser drivers, even higher $\sim 10^5$~T magnetic fields are predicted~\cite{Bukharskii_BullLebedevPhysInst_2023}. 

Although the snail targets are experimentally realized and showed a very good robustness, experimental studies related to application of these targets to proton collimation are yet at the planning stage. Previously, it was found, that depending on the interaction parameters, the structure of the magnetic field in the snail target cavity may be different. However, as the currents are bound to the target surface, at least in the central part the direction of the fields is close to that of the snail axis. Nevertheless, it is not yet clear how much is the internal structure of the field important for the collimation properties of the target and which magnetic field scales are required for collimation in certain proton energetic ranges. 
The main goal of this work is to consider different possible situations and to demonstrate that an efficient collimation of fast protons in magnetic fields, created by short powerful laser beams in a snail target is achievable with modern and perspective laser facilities. \blue{The considered setup for producing high-energy collimated proton beams is illustrated in Fig.~\ref{fig:Bz_prof}, (a). The beam is assumed to be produced via TNSA-mechanism~\cite{Roth_2016} and propagates through the "snail" microcoil, where it interacts with the collimating magnetic field. The resulting beam profile is analyzed at the distances of several tens of mm behind the snail target.}

\section{Magnetic field structure and parameters}\label{sec2}

\begin{figure}[h]
    \centering
    \includegraphics[width=0.9\linewidth]{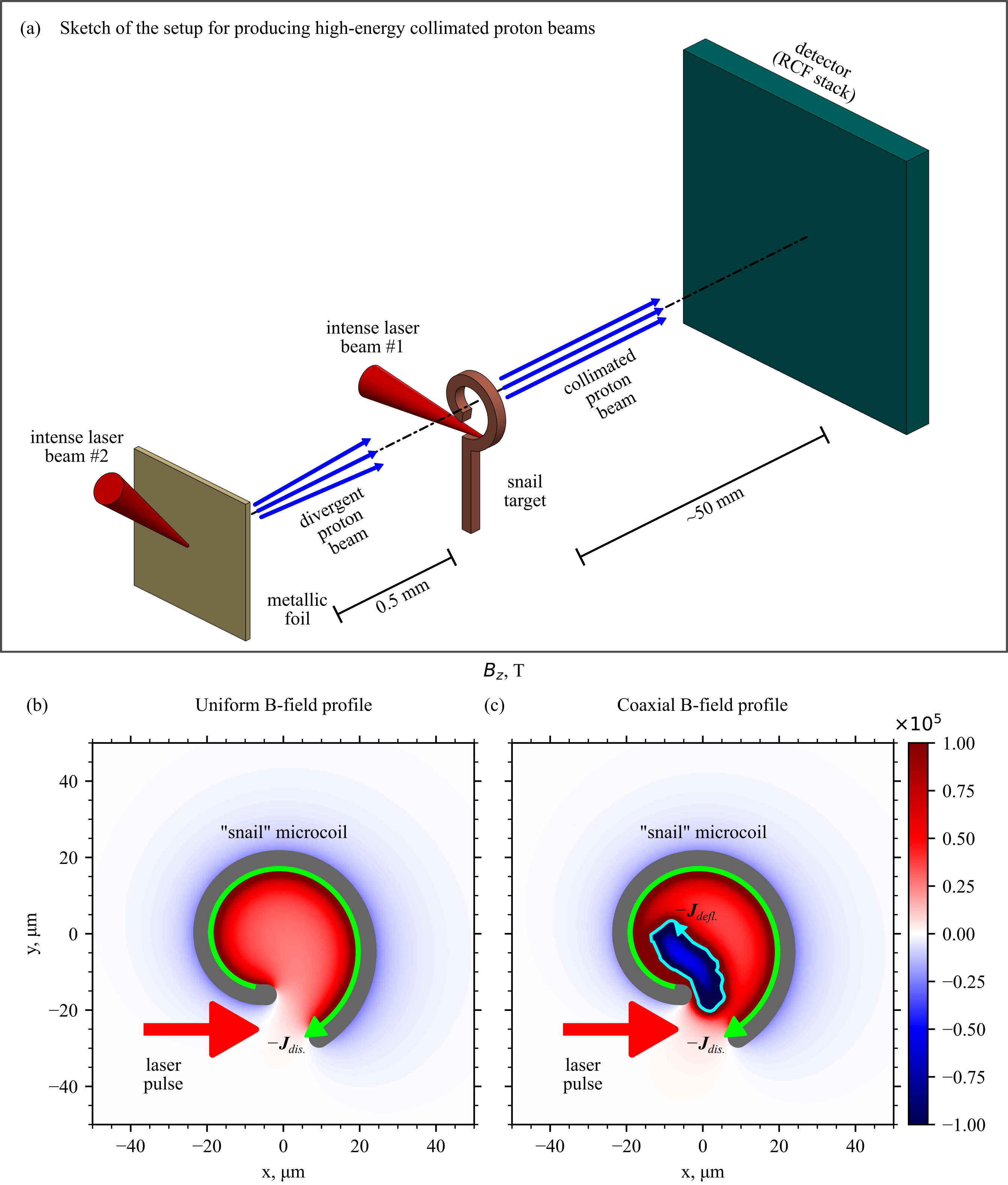}
    \caption{Sketch of the proposed setup for producing high-energy collimated proton beams~(a) and spatial distribution of the transverse magnetic field component $B_z$ inside the snail target: (b)~a simple uniform B-field profile produced by the discharge current along the inner "snail" surface~(light green arrow), (c)~coaxial B-field profile reproducing the one obtained in 2D PIC simulations in~\cite{Bukharskii_BullLebedevPhysInst_2023}, formed by the discharge current of the bulk electrons~(light green arrow) and the current of deflected laser-accelerated electrons~(light blue arrow). Direction of the incident laser pulse used to excite the B-field in "snail" microcoil is schematically shown with the red arrow.}
    \label{fig:Bz_prof}
\end{figure}

In the simulation presented below, the magnetic field profile was defined for a given electric current geometry using the Biot–Savart law. The current geometry was chosen to reproduce either "uniform" or "coaxial" B-field profiles. \blue{By the "uniform" profile we mean hereafter the field profile which is topologically similar to that of the magnetic dipole. The "coaxial" profile was obtained in earlier studies in 2D PIC simulations, when the snail targets are irradiated with very intense but very short laser pulses, see, e.g. in~\cite{Bukharskii_BullLebedevPhysInst_2023}. 
This specific magnetic field structure appears when the current of laser-accelerated electrons is strong enough to magnetize an internal part of the coil. Fig.~\ref{fig:Bz_prof}, (b,c) demonstrates the two magnetic field profiles used in this work.} 
However, in many cases, at least for a ps-duration laser driver, the internal volume with an opposite polarity of the magnetic field was not seen in simulations~\cite{Korneev.etal_GigagaussscaleQuasistaticMagnetic_PRE-2015, Korneev_MagnetizedPlasmaStructures_JPCS-2017, Ehret_PhysRevE_2022}.  

The current geometry providing the magnetic field structure is shown in Fig.~\ref{fig:Bz_prof}, panels (b,c), with arrows. The discharge current is marked with the light green arrow, the current of deflected laser-accelerated electrons is shown with the light-blue arrow, and the red arrow indicates the laser beam propagation. The target profile is shown with the dark gray mask.  Both the uniform and the coaxial magnetic field profiles are shown in the figure. \blue{The electric field is of course also excited in the system as the target is being charged positively under the action of the intense driver \cite{Ehret_PhysRevE_2022}, though its effect in the considered situation is weak, see the discussion below.}

The propagation of high-energy proton beams through the considered electromagnetic structures was modeled using a test-particle approach. \blue{The simulations were performed using a self-developed code written in the high-level programming language Python~\cite{Python}, complemented by the NumPy library~\cite{Numpy} and an open-source JIT-compiler Numba~\cite{Numba} transforming a regular Python code into a fast machine code and allowing parallelization for multiple CPUs.} A beam of protons was sent through the cavity of the "snail" target along the axis~($z$-axis in Fig.~\ref{fig:Bz_prof}, (b,c)). For the propagating particles in the given fields, relativistic equations of motion with the Lorentz force were solved numerically for each particle in the beam using the Boris integration scheme~\cite{Boris_1970}. The fields were assumed to be stationary during the time required for a multi-MeV proton to cross the target region~($100$~MeV protons cross $\simeq 100$~$\upmu$m-long high field region in less than $1$~ps, while the magnetic field changes much slower, on the temporal scale of a few hundreds of ps~\cite{Ehret_PhysRevE_2022, Bukharskii_BullLebedevPhysInst_2023}). 
The interaction between different particles in the proton beam was neglected as space-charge effects are typically low for TNSA-produced proton beams~\cite{Santos_PhysPlasmas_2018}. Proton scattering and energy losses due to collisions with plasma formed in the target cavity were also assumed to be negligible for the considered projectile energies exceeding $10$~MeV scales, taking into account small plasma cloud spatial scales of $\sim 100$~$\upmu$m and characteristic density values of $10^{19}$..$10^{21}$~cm$^{-3}$. The magnetic fields in the target region were calculated on a 3D dimensional grid with 2~$\upmu$m resolution under the assumption that they are produced by the fixed electric currents defined by the contours in Fig.~\ref{fig:Bz_prof}, (b,c). \blue{For this purpose, each of the assumed current paths, i.e. one in the case of the uniform B-field profile (light green in Fig.~\ref{fig:Bz_prof}(b)) or two in the case of the coaxial B-field profile (light green and light blue in Fig.~\ref{fig:Bz_prof}(c)), were divided into $N_I=100$ infinitely thin linear segments of approximately equal lengths. The resulting magnetic field at position $\mathbf{r}$ was found as a sum of contributions from all the $N_I$ segments according to the Biot-Savart law: \[ \mathbf{B} \left( \mathbf{r} \right) = \frac{\mu_0}{4\pi} \sum_{i=1}^{N_I} \frac{I\,d\mathbf{l}_i\times\mathbf{r}_i^{\prime}}{\lvert \mathbf{r}_i^{\prime} \rvert^3}, \] where $I$ is the current value, $d\mathbf{l}_i$ is the vector drawn from the start to the end of the $i$-th linear segment, $\mathbf{r}_i^{\prime}=\mathbf{r}-\mathbf{r}_i$ is the vector from the position $\mathbf{r}_i$ of the $i$-th linear segment to the point $\mathbf{r}$ where the field is calculated, and $\mu_0$ is the magnetic constant.} The current value was $\approx 7 \cdot 10^5$~A, which corresponds to the $10^5$~T scale magnetic fields obtained in 2D PIC simulations in~\cite{Bukharskii_BullLebedevPhysInst_2023}. The electric fields were determined under the assumption that the target is a perfect conductor charged to a certain potential. The latter was set to $200$~kV, similar in scale to the value retrieved from the experimental data in~\cite{Ehret_PhysRevE_2022}. Although, it should be noted that for the considered $\simeq 100$~MeV protons electric fields corresponding to potentials of a few hundred kilovolts have a little effect on trajectories of fast protons -- further modeling of proton beam propagation showed the resulting beam profile remained almost unchanged if the electric potential was varied in $\left( 0-1000 \right)$~kV range. Particles were originated from a point source placed $0.5$~mm away from the "snail" center\blue{, see the related discussion in Section \ref{sec4}}. Their velocities were defined according to their kinetic energy, and their initial angles were set so that the angular distribution of the beam presented a Gaussian profile with the full width at half maximum~(FWHM) of $20^{\circ}$, which corresponds to the half opening angle  $\theta / 2 = 10^{\circ}$. The beam was directed along the "snail" axis and registered at the "detector" plane placed $50$~mm away from the the target plane. \blue{A total of $10^6$ particles were used in the simulation to accumulate sufficient statistics in the detector plane.}
\begin{figure}[h]
    \centering
    \includegraphics[width=0.9\linewidth]{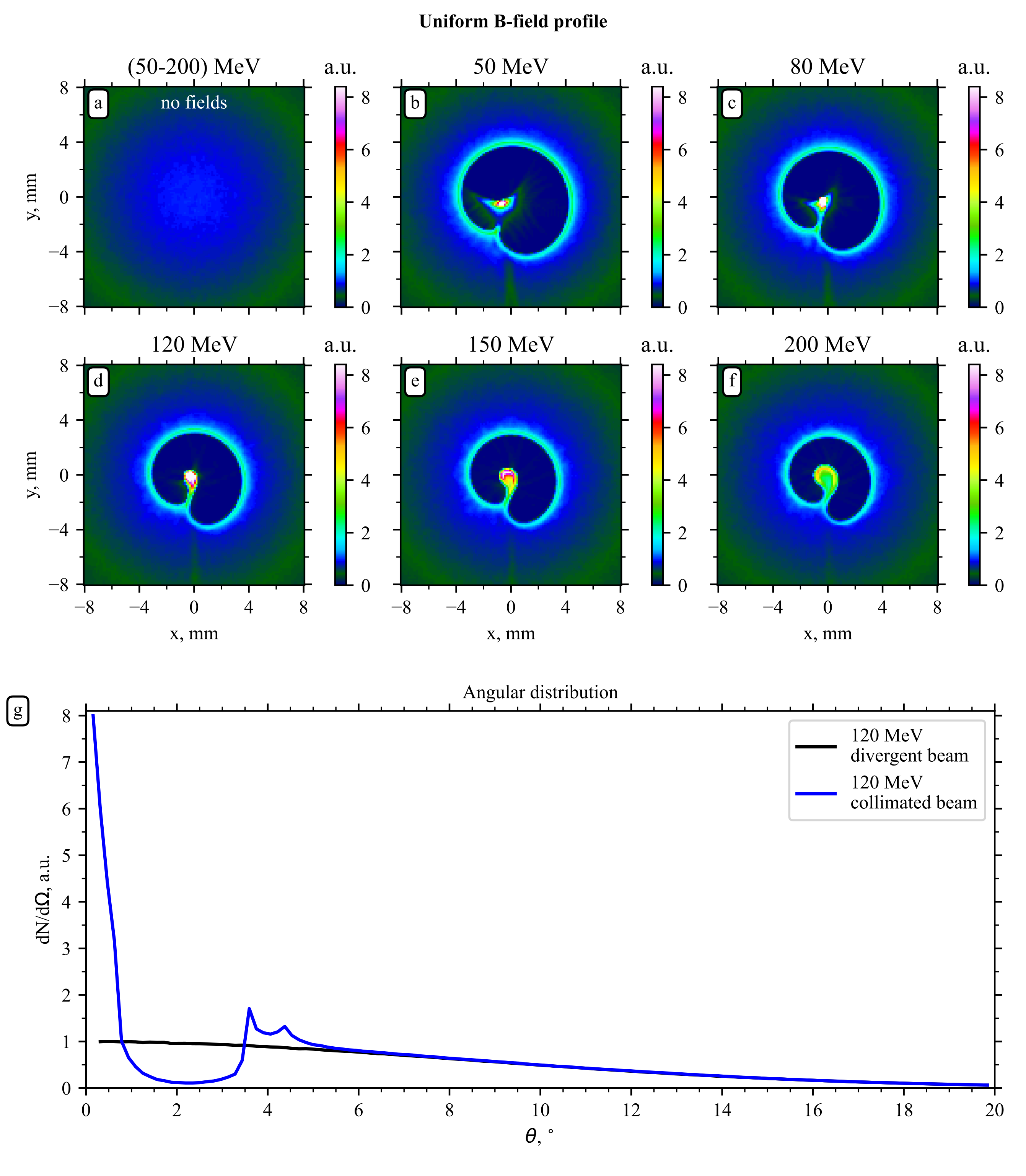}
    \caption{Results of test-particle simulations of proton beam transport through the "snail" cavity for the uniform $\sim 10^5$~T magnetic field and the electric field corresponding to $200$~kV target surface potential: (a)~initial beam profile in the detector plane if no fields are imposed on the beam; (b-f)~beam profiles in the detector plane modified due to the interaction with the B-field around the target for different projectile energies -- $50$~MeV, $80$~MeV, $120$~MeV, $150$~MeV and $200$~MeV, respectively; (g)~angular distribution for the divergent proton beam~(black curve) and collimated beam of $120$~MeV protons~(blue curve).}
    \label{fig:collimation_uni}
\end{figure}
\section{Proton beam collimation in the electromagnetic fields of the "snail" target}\label{sec3}

In the first set of simulations, the proton beam propagation was studied for the uniform B-field profile, see panel~(b) in Fig.~\ref{fig:Bz_prof}. Projectile energies in range $\left( 50-200 \right)$~MeV were considered. The obtained results are presented in Fig.~\ref{fig:collimation_uni}. The presented profiles show that the imposed magnetic field causes a significant redistribution of the particles in the beam. Initially, the beam has a wide profile that occupies most of the detector area, see Fig.~\ref{fig:collimation_uni}, (a). However, when the beam passes through the cavity with the quasi-stationary $10^5$~T scale B-field, protons deflect and form three different regions which can be observed in the detected profiles, panels (b-f). They consist of the "external" region, the "void" region with a bright boundary and the "inner" collimated beam region. The first one, i.e. the "external" region, encompasses all of the area far from the center of the detector plane and it stays almost unmodified relative to the original beam profile, as no significant changes of proton flux are observed in it. This is not surprising as this region corresponds to protons that do not pass through the "snail" cavity, but instead go around it. As the magnetic field strength is considerably greater inside the cavity, it has a little effect on these protons and just slightly deflect them. The second region represents a 'void' with a very low signal. The outer boundary of this region, however, is very bright and has a pronounced caustics. The third and final region is the one inside the 'void' close to its center. This area corresponds to protons that pass through the center of the target close to its axis. These protons are effectively collimated for almost all of the presented cases, see Fig.~\ref{fig:collimation_uni},~(b-f). 

The achieved degree of collimation depends on the projectile energy. For the $\sim 10^5$~T magnetic field produced by the electric current of $7 \cdot 10^5$~A shown in Fig.~\ref{fig:Bz_prof}, (b), the optimal range of energies for which a high degree of collimation is achieved is around $\left( 80-120 \right)$~MeV. If the energy is lower, protons undergo very strong deflection and focus at some point before the detector plane, instead of collimating into a quasi-parallel beam, thus making the resultant pattern at the center of the detector plane irregular and prolonged in certain directions. Protons with higher energies, on the other hand, are under-collimated, and the spot they produce is also irregular. Its shape resembles the shape of the "snail" cavity, while its size is significantly higher than that for protons within the optimal energy range. At the same time, the maximum proton flux for protons outside of the optimal energy range is several times lower. 

In Fig.~\ref{fig:collimation_uni}, (g), angular distributions are presented for the initial divergent beam and the collimated beam of $120$~MeV protons. The presented plot shows the three aforementioned regions the proton beam is divided into: (I)~the "external" region extending from about $6^{\circ}$ upwards; (II)~the 'void' region from $1^{\circ}$ to $4^{\circ}$, enclosed by a bright boundary at $\left( 4.0^{\circ} \pm 0.5^{\circ} \right)$; and (III)~the "inner" collimated spot region at $\left( 0^{\circ}-1^{\circ} \right)$. Due to the redistribution of protons under the action of the magnetic field their flux per unit solid angle along the axis of the beam increases by a factor of $\simeq 8$, while the characteristic divergence angle of the beam decreases from the initial $10^{\circ}$ to the value below $0.5^{\circ}$. This corresponds to more than $100$ times decrease of the solid angle covered by the beam.

\begin{figure}[h]
    \centering
    \includegraphics[width=0.9\linewidth]{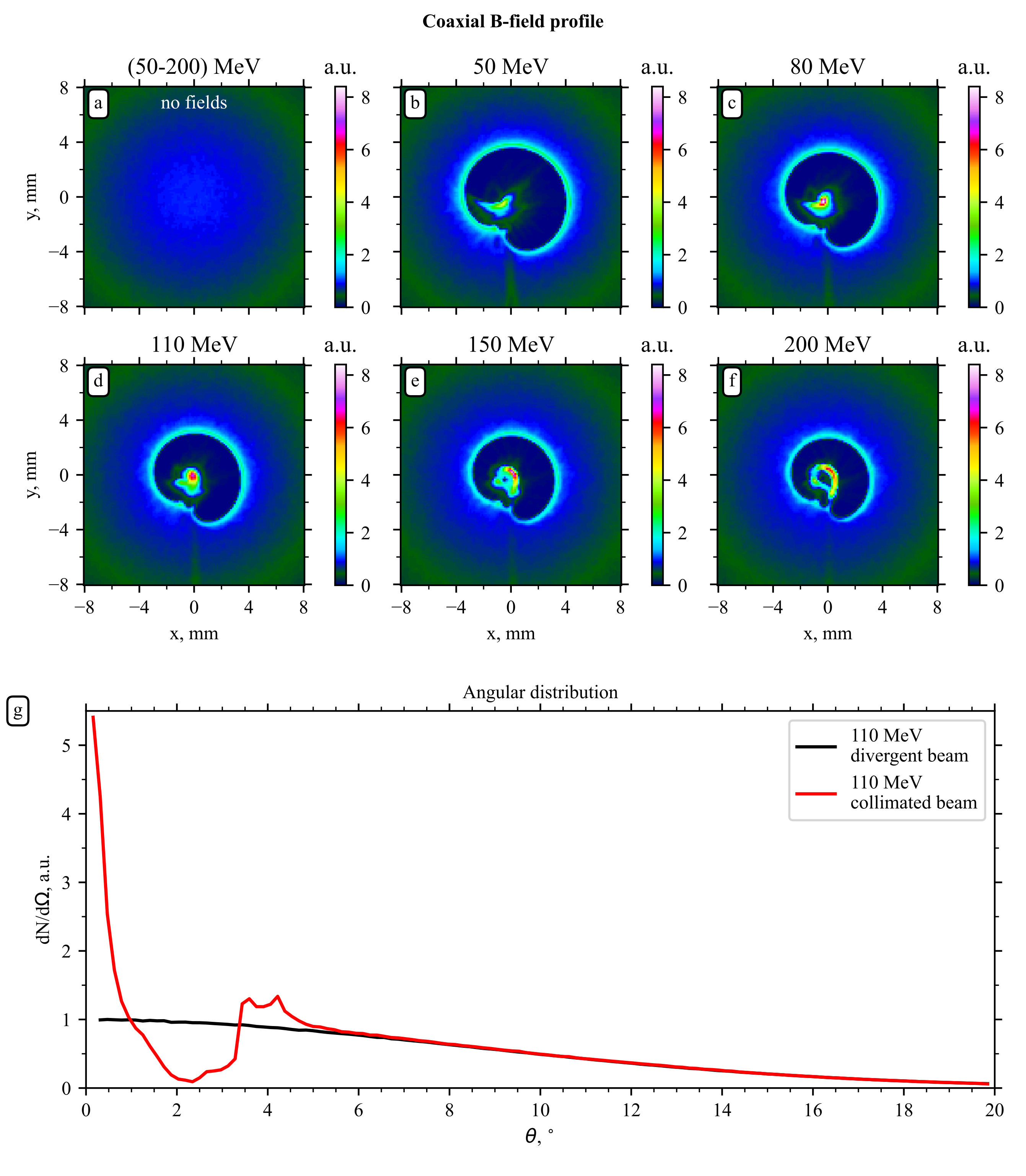}
    \caption{Results of test-particle simulations of proton beam transport through the "snail" cavity for the coaxial $\sim 10^5$~T magnetic field and the electric field corresponding to $200$~kV target surface potential: (a)~initial beam profile in the detector plane if no fields are imposed on the beam; (b-f)~beam profiles in the detector plane modified due to the interaction with the B-field around the target for different projectile energies -- $50$~MeV, $80$~MeV, $110$~MeV, $150$~MeV and $200$~MeV, respectively; (g)~angular distribution for the divergent proton beam~(black curve) and collimated beam of $110$~MeV protons~(red curve).}
    \label{fig:collimation_coax}
\end{figure}

\blue{
The obtained collimation properties of the target may be rather qualitatively described in terms of a magnetic lens approach. Following \cite{Reiser}, in the paraxial approximation, the focal length $f$ for a solenoidal magnetic field lens with the peak magnetic field $B_0$ and the effective field length 
$$
l=\frac{1}{B_0^2}\int_{-\infty}^{\infty}B^2(z)dz,
$$  
is 
\begin{equation}
f=\frac{l}{\Phi \sin\Phi},
\label{focal_length}
\end{equation}
where $\Phi=\omega_B l/2v$, $\omega_B=e B/mc\gamma$ is the gyrofrequency, $v$ is the proton velocity, $B=\int_{-\infty}^{\infty}B(z)dz/l$ is the effective magnetic field, $\gamma=(1-\frac{v^2}{c^2})^{-1/2}$, $c$ is the light velocity, $e$ and $m$ are the proton charge and mass respectively. Using the profiles of the magnetic field, calculated numerically for the "snail" target with the uniform field profile, we find that for the situation considered in Fig. \ref{fig:collimation_uni}, $l\approx 22~\upmu$m, $B\approx21.6$ kT, and then for the proton energies $E=50,~80,~120,~150,~200$ MeV the focal lengthes are $f\approx150,~241,~367,~465,~635~\upmu$m, respectively. 
According to the distributions presented in Fig.~\ref{fig:collimation_uni}, the optimal collimation conditions are realized for the protons with energies of $\sim 120$~MeV. The source of these protons in the simulations was set at the distance of $500~\upmu$m from the target, see Fig.~\ref{fig:Bz_prof}(a), which is about one third greater than the estimated focal length for this energy $f\approx 367~\upmu$m. The probable reason of this difference is the limited applicability of the paraxial approximation in the considered case, as the resulted beam size is of the same order as the coil diameter. It worth mentioning that the thin lens approximation, which would mean expansion $\sin \Phi\approx \Phi$ in \eqref{focal_length}, is not valid as $\Phi$ is not small.}

In the second set of simulations, the proton beam propagates through the target with the coaxial-shape $10^5$~T scale magnetic field, see Fig.~\ref{fig:Bz_prof}(c). The obtained proton beam profiles in the "detector" plane are shown in Fig.~\ref{fig:collimation_coax}, where panels~(b-f) correspond to different proton energies in the range $\left( 50-200 \right)$~MeV. As can be seen, at the optimal and sub-optimal energies, i.e. $\left( 50-120 \right)$~MeV, the profiles do not differ qualitatively from the case of the uniform B-field, see Fig.~\ref{fig:collimation_coax}(b-d), while for higher proton energies the coaxial B-field profile presents some more intricate structures in the proton patterns, which are shown in Fig.~\ref{fig:collimation_coax}(e),(f). The optimal proton energy for which the collimation effect is the highest in the case of the coaxial B-field profile is slightly lower than that for the uniform B-field distribution~($120$~MeV) and equals to $110$~MeV. The shape of the central spot, corresponding to the collimated part of the beam, is very similar, although the peak increase of the proton flux at the center of this spot is somewhat lower and equals to $\simeq 5.5$. The characteristic divergence half-angle of the collimated beam is $\approx0.5^{\circ}$, similar to that for the uniform B-field distribution.

\begin{figure}[h]
    \centering
    \includegraphics[width=0.9\linewidth]{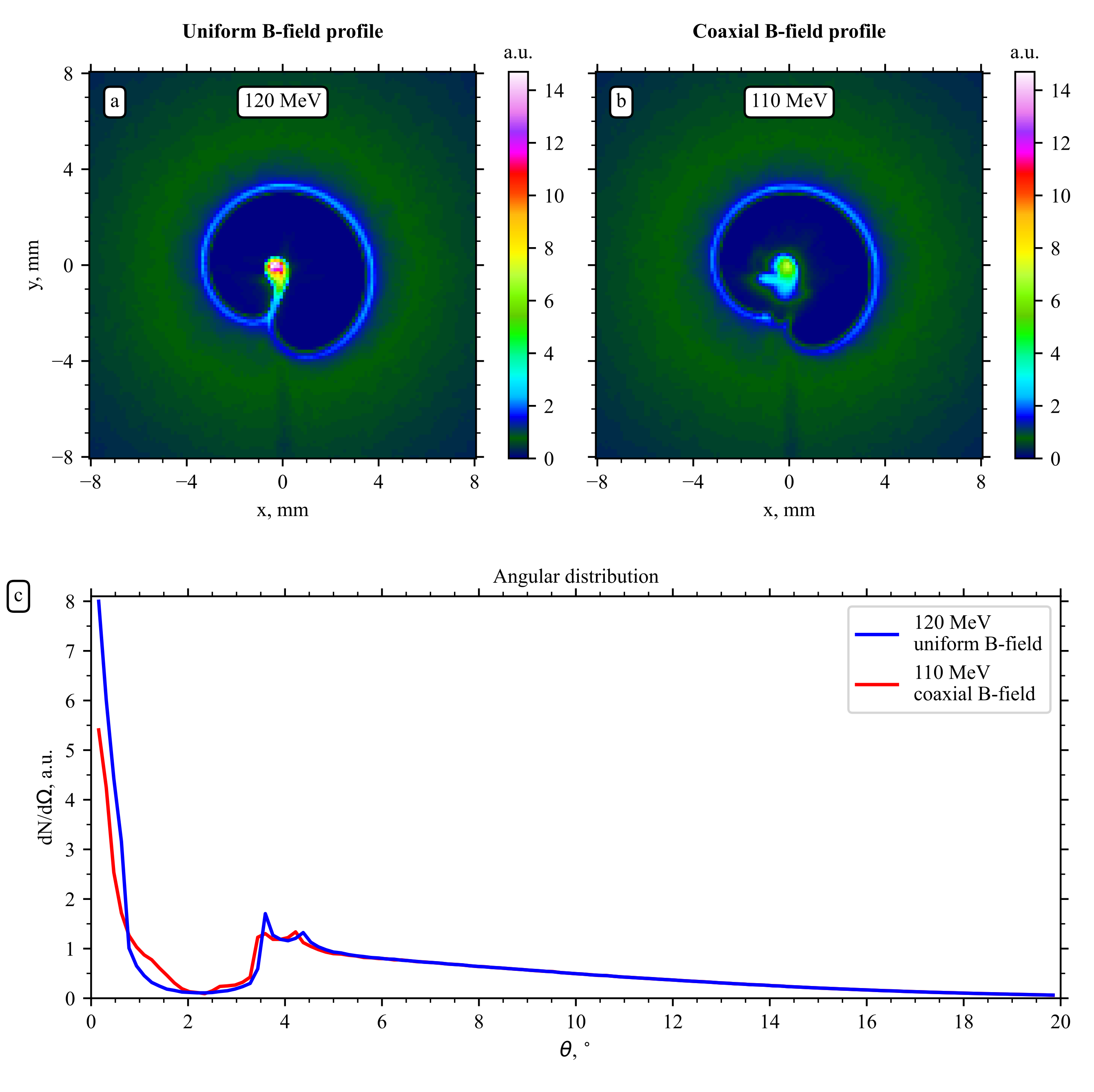}
    \caption{Comparison of the results of test-particle simulations of proton beam transport through the "snail" cavity for different B-field profiles: (a,b)~beam profiles in the detector plane for the uniform and coaxial B-field profiles, with projectile energies of $120$~MeV and $110$~MeV, respectively; (c)~angular distribution for the beam of $120$~MeV protons collimated by the uniform B-field~(blue curve) and $110$~MeV protons collimated by the coaxial B-field~(red curve).}
    \label{fig:collimation_comp}
\end{figure}
Fig.~\ref{fig:collimation_comp} shows an explicit comparison of the resulting proton beam profiles for the uniform and coaxial collimating field structures. Panels (a,b), where the two-dimensional distributions are presented, illustrate that the coaxial field structure is slightly inferior to the uniform B-field distribution of the same strength -- with the same color scale, the proton beam intensity in the focal spot appears to be less bright. The spot itself is also somewhat larger in panel~(b), and proton density does not drop as steeply at its edges as for the uniform B-field case shown in panel~(a). The analysis of the angular distributions also show that less particles propagate along the target axis during interaction with the coaxial $10^5$~T B-field in comparison to the uniform $10^5$~T B-field profile. However, the difference amounts to only a few tens of percent, and thus, both the coaxial and the uniform B-field profiles in the considered range of parameters can be used effectively for the proton beam collimation. 

\section{Discussion}\label{sec4}

\begin{figure}[h]
    \centering
    \includegraphics[width=0.9\linewidth]{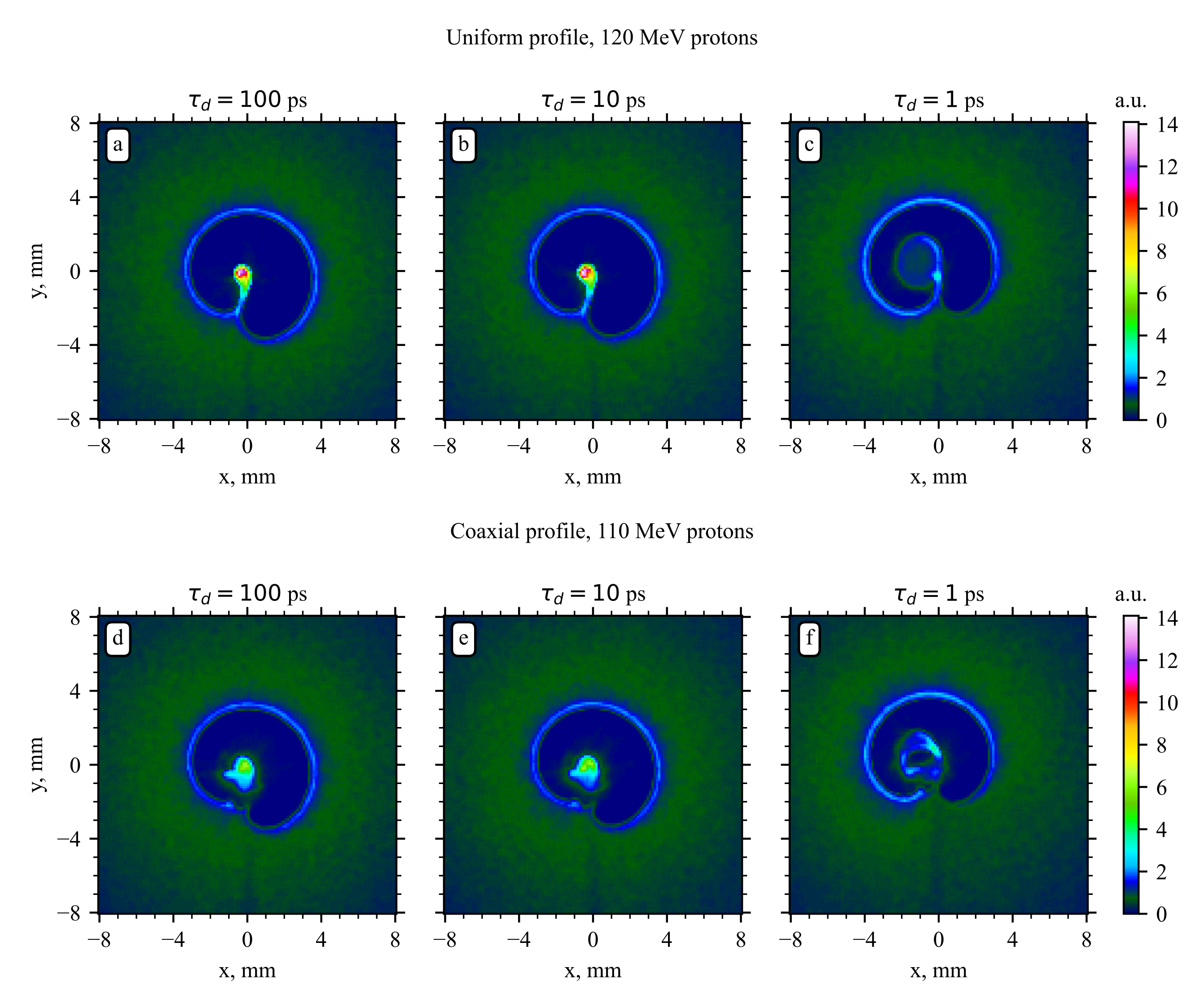}
    \caption{Results of test-particle simulations of proton beam transport through the "snail" cavity for different B-field profiles, uniform~(top row) and coaxial~(bottom row), with the inductive electric field caused by the time-varying magnetic field taken into account. Panels~(a,d) correspond to the magnetic field decay time $\tau_d = 100$~ps, panels~(b,e) -- to $\tau_d = 10$~ps and panels~(c,f) -- to $\tau_d=1$~ps.}
    \label{fig:collimation_with_inductive_field}
\end{figure}

\blue{The calculations described in Sec.~\ref{sec3} were performed under the assumption that the magnetic field induced around the target remains static and thus does not excite vortex electric field. To justify the validity of such assumption for the considered problem, an additional set of simulations was performed to study the effect of the inductive electric field on particle collimation. This field was calculated as
\begin{equation}
    \mathbf{E_i} = - \frac{1}{c} \frac{\partial \mathbf{A}}{\partial t},
    \label{eq:inductive_electric_field_from_vector_potential}
\end{equation}
where $\mathbf{A}$ denotes the retarded vector potential:
\begin{equation}
    \mathbf{A} \left( \mathbf{r}, t \right) = \frac{1}{c} \int_{\Omega} \frac{\mathbf{J} \left( \mathbf{r^{\prime}}, t^{\prime} \right)}{\left| \mathbf{r} - \mathbf{r^{\prime}} \right|} d^3 \mathbf{r^{\prime}}.
    \label{eq:retarded_vector_potential}
\end{equation}
The current in Eq.~(\ref{eq:retarded_vector_potential}) was constant along each of the selected paths, see Fig.~\ref{fig:Bz_prof}, panels~(b,c), and decayed linearly in time from $\sim 7 \cdot 10^5$~A~(corresponding magnetic field is $B\sim 10^5$~T), down to zero. To demonstrate the effect, different decay times were considered, $\tau_d = 100$~ps, $\tau_d = 10$~ps and $\tau_d = 1$~ps, where the value $\tau_d = 100$~ps corresponds to the one retrieved from experimental data for "snail" targets in~\cite{Ehret_PhysRevE_2022}. Resulting beam profiles are shown in Fig.~\ref{fig:collimation_with_inductive_field}.}

\blue{The inductive electric field induced by the decaying annular current is also annular in the target cavity region. Its main component is $E_{\phi}$ where $\phi$ denotes the azimuthal angle measured in the plane perpendicular to the beam propagation axis. Such electric field changes projectile's azimuthal velocity component, which in turn causes a change of the radial component velocity component, as $\mathbf{v}_\phi \times \mathbf{B}$ force is directed radially inside the "snail" cavity. From the obtained distributions for the uniform B-field profile it follows that if the B-field changes on the time-scale $\sim \left( 10-100 \right)$~ps, the described effect is insignificant, see panels (a) and (b) in Fig.~\ref{fig:collimation_with_inductive_field}, where the beam profiles change only very slightly in comparison to the those presented in Sec.~\ref{sec3}, with a decrease of peak proton density at the center of the collimated beam not exceeding a few percent.  For $\tau_d = 1$~ps, however, the inductive electric field strength appears to be strong enough to cause a significant redistribution of protons and reduce the collimation efficiency. This case is shown in panel~(c) of Fig.~\ref{fig:collimation_with_inductive_field}. The same applies to the beam profiles obtained for the coaxial B-field structure, shown in panels~(d-f). As for the uniform B-field profile, the inductive electric field impairs collimation only for $\tau_d = 1$~ps, see panel~(f) in Fig.~\ref{fig:collimation_with_inductive_field}. Therefore, for the significant inductive electric field contribution in the considered scheme, the magnetic field has to change on $\sim1$~ps time-scale, which is much faster than the B-field decay rate measured experimentally~\cite{Ehret_PhysRevE_2022} and observed in simulations \cite{Bukharskii_BullLebedevPhysInst_2023}. Thus, in the considered range of parameters, the inductive fields do not play a significant role in the beam collimation. The static electric fields for realistic target potentials up to $1$~MV also appear to be weak to cause a significant deflection of $\simeq 100$~MeV protons, as it was verified by an auxiliary set of simulations. Thus, we can conclude that for the considered high energies of protons and strong values of the magnetic field, the collimation effect is mainly defined by the quasistationary magnetic field induced around the target.}

\begin{figure}[h]
    \centering
    \includegraphics[width=0.9\linewidth]{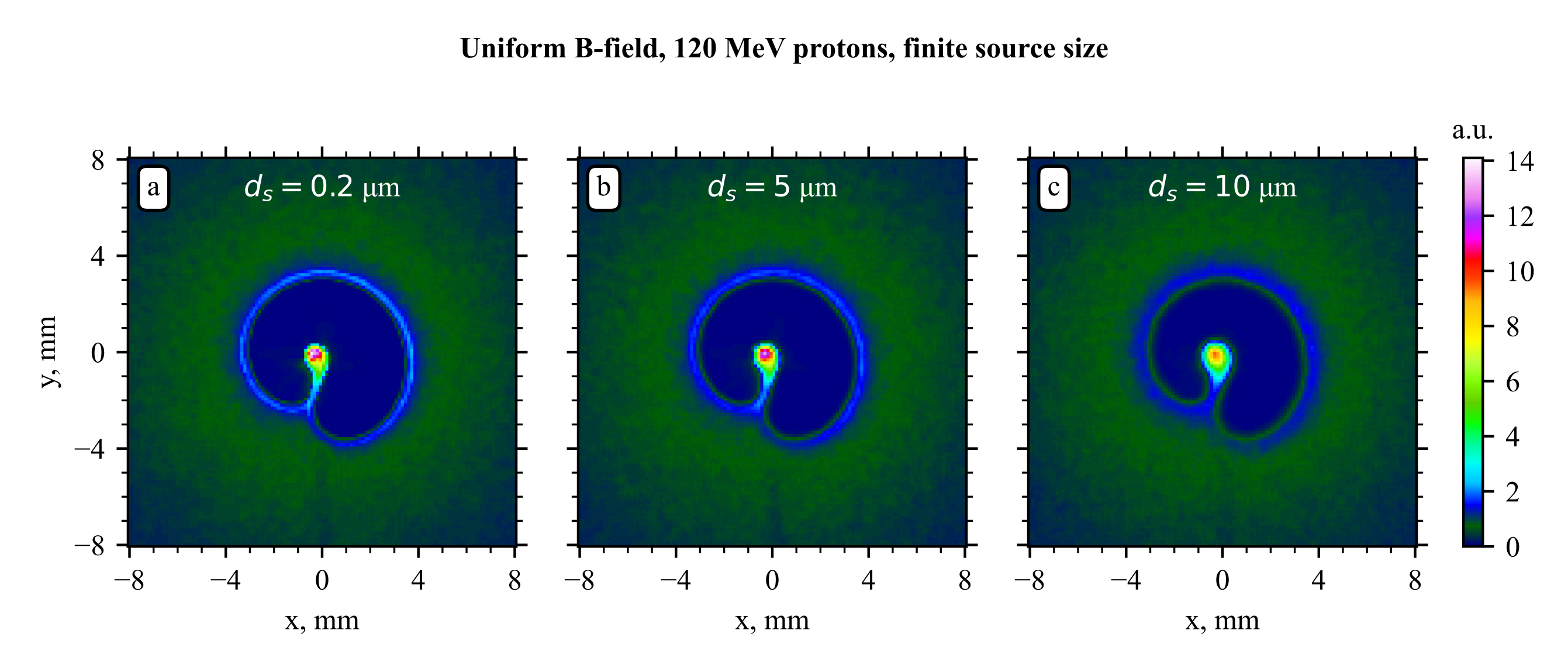}
    \caption{Results of test-particle simulations of proton beam transport through the "snail" cavity for the uniform B-field, $120$~MeV proton energy and different source sizes $d_s$: (a)~point-like source with $d_s = 0.2$~$\upmu$m, (b,c)~finite-size sources with $d_s=5$~$\upmu$m and $d_s=10$~$\upmu$m, respectively.}
    \label{fig:collimation_finite_source}
\end{figure}

\blue{Another important point worth discussing is the used model of the proton beam -- in Sec.~\ref{sec3} a simple ideal case was considered, where a monoenergetic proton beam is emitted from a point-like source, while in reality the source has finite dimensions and some energy spread.}
\blue{Assuming that the proton beam is produced via the TNSA mechanism during an ultrashort laser pulse irradiation of a thin metallic foil. The imaging properties of such a source, despite relatively large transverse size of the emitting region of $\sim 100-200$~$\upmu$m, can be reproduced by assuming a much smaller source shifted by $\sim 100$~$\upmu$m from the foil surface~\cite{Borghesi_PRL_2004}. The effective size of such a source can be as small as $\simeq 5-10$~$\upmu$m~\cite{Borghesi_PRL_2004, Kugland_RSI_2012}. As discussed in~\cite{Kugland_RSI_2012}, finite source size may lead to blurring of proton images, i.e. beam profiles, and can be neglected if the source size $d_s$ is much lower than the probed field size $d$: $d_s \ll d$. Here, assuming $d_s \simeq 5-10$~$\upmu$m the condition $d_s \ll d$ is satisfied as $d \simeq 50$~$\upmu$m, and blurring related to the finite source size should be insignificant, which was verified by an auxiliary set of simulations. There, the initial transverse proton positions were chosen randomly inside circles with different diameters, $d_s = 0.2$~$\upmu$m, which corresponds to an almost point-like source, as well as $d_s = 5$~$\upmu$m and $d_s = 10$~$\upmu$m which describe finite-size sources. Results of the performed simulations for the uniform B-field profile are shown in Fig.~\ref{fig:collimation_finite_source}. As can be seen, blurring becomes significant only for $d_s = 10$~$\upmu$m, see Fig.~\ref{fig:collimation_finite_source}(c). In this case, the peak proton density drops by $\approx 35$~\% in comparison to the point-like source case, shown in panel~(a) of the same figure. For $d_s=5$~$\upmu$m, the peak proton density decrease due to the blurring amounts to a modest $10$~\%. Similar results were obtained assuming the coaxial B-field profile. Thus, for the considered scheme the point source assumption remains viable for $d_s \lesssim 5$~$\upmu$m as the transverse source size in this case is much smaller than the collimating field transverse size, and the collimation efficiency remains almost unaffected.}

\begin{figure}[h]
    \centering
    \includegraphics[width=0.9\linewidth]{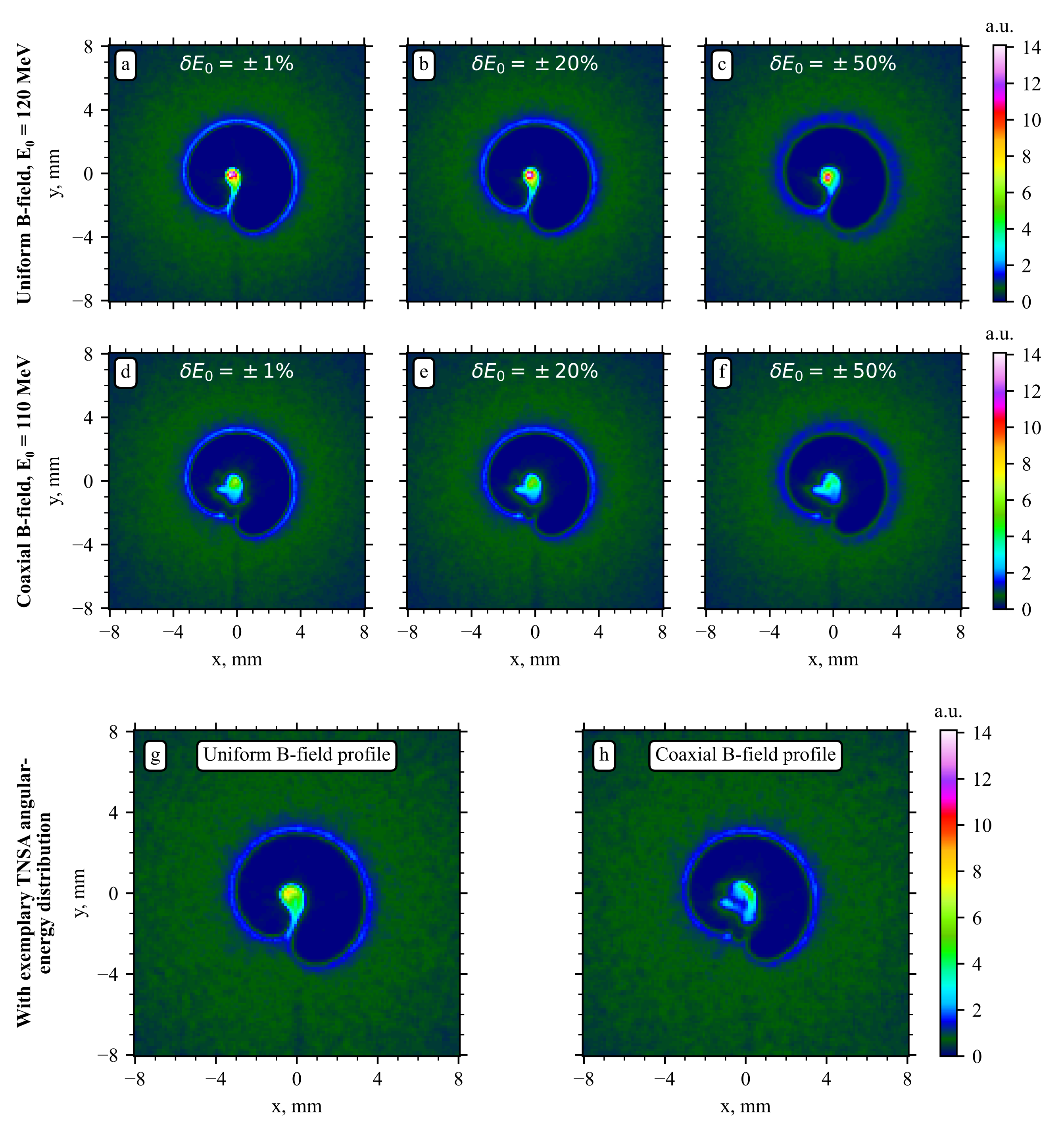}
    \caption{Results of test-particle simulations of a non-monoenergetic proton beam transport through the "snail" cavity: (a-c)~$120$~MeV protons in the uniform B-field with different values of energy spread, $\delta E_0 = 1$~\%, $\delta E_0 = 20$~\% and $\delta E_0 = 50$~\%, respectively; (d-f)~$110$~MeV protons in the coaxial B-field with the same values of the energy spread, $\delta E_0 = 1$~\%, $\delta E_0 = 20$~\% and $\delta E_0 = 50$~\%, respectively; (g-h)~beam profiles for the uniform and coaxial B-field profiles assuming an exemplary angular-energy distribution for the TNSA-mechanism.}
    \label{fig:collimation_nonmonoenergetic}
\end{figure}

\blue{In terms of the energy spectrum, TNSA-produced protons are usually characterized by a wide energy spread~\cite{Roth_2016}. Although, there have been a number of reports of quasi-monoenergetic ion beams produced via TNSA-mechanism using specifically-designed and prepared targets~\cite{Hegelich_Nature_2006, Schwoerer_Nature_2006, TerAvetisyan_PRL_2006} -- there, the energy spread of accelerated ions did not exceed $20$~\%. Despite relative difficulties in obtaining monoenergetic ion beams with TNSA, it should be mentioned that for practical applications involving interaction of the TNSA-produced proton beams with thick targets the target itself plays a role on an energy filter due to the presence of well-localized region of ion losses in matter, the so-called Bragg peak. In this case, only a small fraction of the whole beam energy from a certain part of the spectrum, is deposited in the target at a certain depth. This allows for the used description of the monoenergetic proton beams in Sec.~\ref{sec3}. Nevertheless, it is worth studying how the energy spread effects the collimation efficiency. The study was performed both for the uniform and coaxial B-field profiles, assuming three different values of the energy spread -- $\delta E_0 = \pm 1$~\%, i.e. almost a monoenergetic beam, as well as $\delta E_0 = \pm 20$~\% and $\delta E_0 = \pm 50$~\%. Results are summarized in Fig.~\ref{fig:collimation_nonmonoenergetic}, panels~(a-f).}

\blue{From the obtained distributions it follows that the decrease of the collimation efficiency is relatively modest, even for $\delta E_0 = 50$~\% -- the collimated spot retains almost the same shape as in the case of a quasi-monoenergetic beam shown in Fig.~\ref{fig:collimation_nonmonoenergetic}(a,d), but the peak proton density at the center of this spot slightly drops. For $\delta E_0 = 20$~\% which has already been successfully achieved in several experiments on laser-driven ion acceleration~\cite{Hegelich_Nature_2006, Schwoerer_Nature_2006, TerAvetisyan_PRL_2006}, this drop amounts to only $\simeq \left( 6-12 \right)$~\%, see Fig.~\ref{fig:collimation_nonmonoenergetic}(b,e). For higher $\delta E_0$ values, the decrease of the peak proton density depends on the B-field structure -- for the uniform profile, it still remains relatively small, about $12$~\%, see Fig.~\ref{fig:collimation_nonmonoenergetic}(c), while for the coaxial profile it amounts to $\approx 30$~\%, see Fig.~\ref{fig:collimation_nonmonoenergetic}(f).}

\blue{In addition, to provide some practical example, we have also considered collimation of a proton beam with angular and energy distributions mimicking those of a real TNSA-based source. Its energy spectrum was defined by an exponential function~\cite{Roth_2016}: $\frac{dN}{dE} \sim \frac{1}{\sqrt{E}} \exp{\left( 
- \sqrt{\frac{2E}{k_BT_{hot}}} \right)}$, with $k_BT_{hot} \approx 125$~MeV being the hot electron temperature estimated using the ponderomotive scaling: $k_BT_{hot} = m_e c^2 \left( \sqrt{1 + \frac{I_0\,\left[ \text{W/cm$^2$}\right]\,\lambda^2\,\left[ \text{$\upmu$m$^2$} \right]}{1.37 \cdot 10^{18}}} - 1 \right)$, where $I_0 = 10^{23}$~W/cm$^2$ and $\lambda = 0.910$~$\upmu$m correspond to the laser pulse parameters, i.e. peak intensity and wavelength, of the XCELS facility~\cite{Kostyukov_BullLebedevPhysInst_2023}. The spectrum extended from $1$~MeV to $250$~MeV. To imitate the energy dependence of the opening angle, the angular distribution of the source was also modified -- it still presented a Gaussian, but its width decreased with energy as an inverted parabolic function~\cite{Roth_2016}: $\theta = \theta_0 - \left( \theta_0 - \theta_{min} \right) \cdot \frac{E^2}{E_{max}^2}$, where $\theta_0/2=20^{\circ}$ defined the divergence for zero energy, and $\theta_{min}/2=5^{\circ}$ defined the minimum divergence for the most energetic part of the spectrum, with $E\simeq E_{max} = 250$~MeV. Results of the performed simulations for the uniform and coaxial B-field profiles are presented in Fig.~\ref{fig:collimation_nonmonoenergetic}(g,h). They indicate that the collimation effect for a beam with a TNSA-like angular-energy distribution still persists, though such beam as a whole collimates less efficiently in comparison to the quasi-monoenergetic beam with a small energy spread shown in panels Fig.~\ref{fig:collimation_nonmonoenergetic}(a,d). Both the shape of the central spot and the peak proton density appear to be affected -- for the uniform B-field profile, see Fig.~\ref{fig:collimation_nonmonoenergetic}(g), the former increases by $\simeq \left( 2-3 \right)$ times while the latter drops by $1.8$~times relative to the values obtained for a quasi-monoenergetic beam with $\delta E_0 = 1$~\%. Interestingly, in the case of the coaxial B-field profile, see Fig.~\ref{fig:collimation_nonmonoenergetic}(h), for a realistic TNSA-like angular-energy distribution, a drop of the peak proton density is much lower, $\sim \left( 10-20 \right)$~\%, though the collimated spot seems to be much more irregular.}

\begin{figure}[h]
    \centering
    \includegraphics[width=0.9\linewidth]{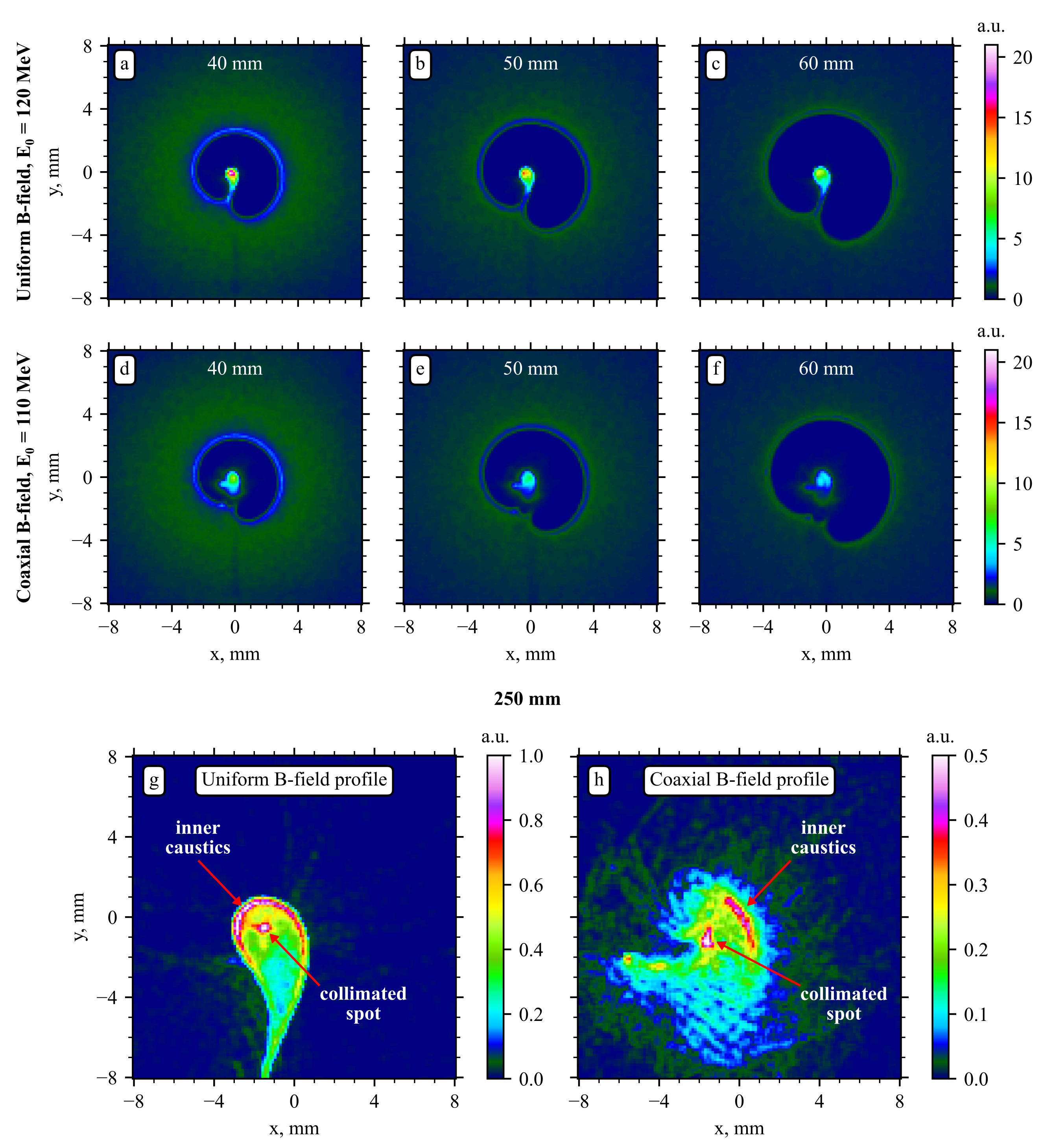}
    \caption{Results of test-particle simulations of proton beam transport through the magnetized "snail" target: (a-c)~beam profiles for $120$~MeV protons and for the uniform B-field at different distances from the target plane -- $40$~mm, $50$~mm and $60$~mm, respectively; (d-f)~beam profiles for $110$~MeV protons  and the coaxial B-field at the same three distances from the target plane -- $40$~mm, $50$~mm and $60$~mm, respectively; (g)~the beam profile for $120$~MeV protons and the uniform B-field at $250$~mm distance from the target plane; (h)~the beam profile for $110$~MeV protons and the coaxial B-field at $250$~mm distance from the target plane.}
    \label{fig:collimation_longitudinal}
\end{figure}

\blue{Finally, it is also worth discussing how the beam profile changes as it propagates along its axis. Above, distributions in a single plane at a fixed distance of $50$~mm from the target plane were provided. To illustrate the beam evolution as is propagates in the longitudinal direction, transverse beam profiles were compared at different distances between the detector and the target plane -- $40$~mm, $50$~mm and $60$~mm~(i.e. in $\pm 10$~mm distance range from the plane where the profiles had been calculated previously) and additionally at $250$~mm~(i.e. $5$ times farther from the target than before). The results are shown in Fig.~\ref{fig:collimation_longitudinal}.}

\blue{Examination of the subsequent transverse cuts at $\left( 50 \pm 10 \right)$~mm both for the uniform~(Fig.~\ref{fig:collimation_longitudinal}(a-c)) and the coaxial B-field profiles~(Fig.~\ref{fig:collimation_longitudinal}(d-f)) indicates that the collimated beam in not completely parallel and retains some residual divergence, as was already illustrated, e.g. in panels~(c) of Figs.~\ref{fig:collimation_uni} and \ref{fig:collimation_coax}. This results in a linear decrease of the central spot size with distance, so that at the $250$~mm distance it appears to be about $5$~times larger than at the $50$~mm distance, while the peak proton concentration drops as the distance squared, so that $250$~mm away from the target it is about twenty 
times lower than at the $50$~mm distance, see Fig.~\ref{fig:collimation_longitudinal}(a-f) and Fig.~\ref{fig:collimation_longitudinal}(g,h). This can be in part a consequence of an imperfectly symmetric field structure, i.e. asymmetry of the field distribution relative to the beam propagation axis. Yet, the imposed magnetic field has a strong positive effect on the peak proton density -- in the central spot corresponding to protons that pass through the target cavity for all distances it increases by $\simeq \left( 7-14 \right)$~times, depending on the field structure. At $250$~mm distance from the target, the inner structure of the central spot visible in panels~(a-f) becomes resolved, see panels (g,h) of Fig.~\ref{fig:collimation_longitudinal}. Interestingly, it may also have additional, "inner" caustics, which are better visible in the case of the uniform field, see panel~(g). A small spot formed by collimated protons is visible close to the center of the profile in both panels~(g) and (h). At a large distance it can be seen that this most collimated proton fraction slightly changes its propagation direction -- the spot is shifted from $\left(0, 0\right)$ position at the detector plane, by the shift is small, with the deflection angle relative to the propagation axis of $\simeq \left( 0.3^{\circ}-0.4^{\circ} \right)$.} 

\section{Conclusion}\label{sec5}

In this work, collimation of $\simeq 100$~MeV protons by electromagnetic structures with $10^5$~T scale magnetic fields induced inside a "snail" target was studied using test-particle simulations. Two qualitatively different magnetic field profiles were considered -- a simple uniform distribution formed solely by the discharge current flowing along the inner "snail" surface, and a more complex coaxial structure, where in addition to the discharge current, currents of laser-accelerated electrons deflected in the forming magnetic field also provide a significant contribution to the resulting B-field profile. The both considered B-field profiles were shown to be suitable for collimation of high-energy protons. The uniform structure, according to the results of the presented numerical modeling, can be used to collimate $120$~MeV protons providing $8$~times increase of the proton flux at the center of the collimated beam. The coaxial B-field profile was shown to be slightly inferior to the uniform structure -- according to the performed simulations, it can be used for collimation of $110$~MeV protons providing $5.5$~times increase of their concentration. Despite about a $30$~\% difference in the maximum achieved proton flux, both the uniform and the coaxial B-field profiles were shown to decrease the beam divergence by more than a factor of $10$, from $10^{\circ}$ to $\approx 0.5^{\circ}$~(FWHM). This corresponds to $\geq 100$-fold decrease of the solid angle covered by the proton beam. 

\blue{The collimation effect for the scheme considered here is attributed mainly to the action of the $\sim 10^5$~T scale magnetic field, while static and inductive electric fields for realistic parameters do not have a significant impact on the collimation efficiency. The effect of finite source size appears to be negligible, as long as it stays at least an order of magnitude lower than the size of the collimating cavity. Here, in order not to impair the collimation efficiency, the effective source size has to be $\lesssim 5$~$\upmu$m. Although, the considered system can be used to collimate TNSA-produced beams with a wide energy spectrum, it is better suited for the collimation of quasi-monoenergetic beams with energy spread not exceeding $\simeq 20$~\%. However, collimation of beams with wide energy spectrum can be improved by fine-tuning of the B-field decay time -- it can be optimized in such a way, that the field's decay rate matches the time of flight for protons with a particular energy to provide their efficient collimation, turning the considered system into an achromatic magnetic lens. In this case, protons with different energies are assumed to approach the magnetic field region in different times, the lower the energy the greater this time, depending on the initial proton source position. For slower protons the magnetic field needed for their efficient collimation is lesser, and the suitable values may be in principle adjusted using different materials, thicknesses or structures of the targets.} 

\blue{It was found that although, the beam retains some residual divergence $\simeq 1^{\circ}$, probably as a consequence of its non-axisymmetric structure, the magnetic field of the snail target still greatly increases proton concentration at the center of the detector plane -- a $\simeq \left( 7-14 \right)$ increase is observed at all distances from the target plate relative to the initial divergent beam case.} 
The proposed method can be used at modern and perspective petawatt and multi-petawatt laser facilities~(e.g.~\cite{ Danson_HPLSE_2019,Kostyukov_BullLebedevPhysInst_2023}) as a part of an integrated all-optical acceleration and guiding setup. In such a scheme, e.g. the TNSA-produced protons created by one laser pulse may pass the optically-driven electromagnetic structure of the considered "snail"-type target. Such a source of fast protons is expected to have an exceptional brightness, low divergence and high energy of accelerated particles, and it would be of interest for fundamental studies and prospective applications.

\section{Acknowledgements}

The work was funded by the Russian Science Foundation under Grant No. 24-22-00402. The authors acknowledge the NRNU MEPhI High-Performance Computing Center and the Joint Supercomputer Center of RAS.

\bibliography{sn-bibliography}

\begin{thebibliography}{52}%
\makeatletter
\providecommand \@ifxundefined [1]{%
 \@ifx{#1\undefined}
}%
\providecommand \@ifnum [1]{%
 \ifnum #1\expandafter \@firstoftwo
 \else \expandafter \@secondoftwo
 \fi
}%
\providecommand \@ifx [1]{%
 \ifx #1\expandafter \@firstoftwo
 \else \expandafter \@secondoftwo
 \fi
}%
\providecommand \natexlab [1]{#1}%
\providecommand \enquote  [1]{``#1''}%
\providecommand \bibnamefont  [1]{#1}%
\providecommand \bibfnamefont [1]{#1}%
\providecommand \citenamefont [1]{#1}%
\providecommand \href@noop [0]{\@secondoftwo}%
\providecommand \href [0]{\begingroup \@sanitize@url \@href}%
\providecommand \@href[1]{\@@startlink{#1}\@@href}%
\providecommand \@@href[1]{\endgroup#1\@@endlink}%
\providecommand \@sanitize@url [0]{\catcode `\\12\catcode `\$12\catcode
  `\&12\catcode `\#12\catcode `\^12\catcode `\_12\catcode `\%12\relax}%
\providecommand \@@startlink[1]{}%
\providecommand \@@endlink[0]{}%
\providecommand \url  [0]{\begingroup\@sanitize@url \@url }%
\providecommand \@url [1]{\endgroup\@href {#1}{\urlprefix }}%
\providecommand \urlprefix  [0]{URL }%
\providecommand \Eprint [0]{\href }%
\providecommand \doibase [0]{http://dx.doi.org/}%
\providecommand \selectlanguage [0]{\@gobble}%
\providecommand \bibinfo  [0]{\@secondoftwo}%
\providecommand \bibfield  [0]{\@secondoftwo}%
\providecommand \translation [1]{[#1]}%
\providecommand \BibitemOpen [0]{}%
\providecommand \bibitemStop [0]{}%
\providecommand \bibitemNoStop [0]{.\EOS\space}%
\providecommand \EOS [0]{\spacefactor3000\relax}%
\providecommand \BibitemShut  [1]{\csname bibitem#1\endcsname}%
\let\auto@bib@innerbib\@empty
\bibitem [{\citenamefont {Yoon}\ \emph {et~al.}(2021)\citenamefont {Yoon},
  \citenamefont {Kim}, \citenamefont {Choi}, \citenamefont {Sung},
  \citenamefont {Lee}, \citenamefont {Lee},\ and\ \citenamefont
  {Nam}}]{Yoon_Optica_2021}%
  \BibitemOpen
  \bibfield  {author} {\bibinfo {author} {\bibfnamefont {J.~W.}\ \bibnamefont
  {Yoon}}, \bibinfo {author} {\bibfnamefont {Y.~G.}\ \bibnamefont {Kim}},
  \bibinfo {author} {\bibfnamefont {I.~W.}\ \bibnamefont {Choi}}, \bibinfo
  {author} {\bibfnamefont {J.~H.}\ \bibnamefont {Sung}}, \bibinfo {author}
  {\bibfnamefont {H.~W.}\ \bibnamefont {Lee}}, \bibinfo {author} {\bibfnamefont
  {S.~K.}\ \bibnamefont {Lee}}, \ and\ \bibinfo {author} {\bibfnamefont
  {C.~H.}\ \bibnamefont {Nam}},\ }\bibfield  {title} {\enquote {\bibinfo
  {title} {Realization of laser intensity over $10^{23}$~w/cm$^2$},}\ }\href
  {\doibase 10.1364/OPTICA.420520} {\bibfield  {journal} {\bibinfo  {journal}
  {Optica}\ }\textbf {\bibinfo {volume} {8}},\ \bibinfo {pages} {630--635}
  (\bibinfo {year} {2021})}\BibitemShut {NoStop}%
\bibitem [{\citenamefont {Mourou}, \citenamefont {Tajima},\ and\ \citenamefont
  {Bulanov}(2006)}]{Mourou_RevModPhys_2006}%
  \BibitemOpen
  \bibfield  {author} {\bibinfo {author} {\bibfnamefont {G.~A.}\ \bibnamefont
  {Mourou}}, \bibinfo {author} {\bibfnamefont {T.}~\bibnamefont {Tajima}}, \
  and\ \bibinfo {author} {\bibfnamefont {S.~V.}\ \bibnamefont {Bulanov}},\
  }\bibfield  {title} {\enquote {\bibinfo {title} {Optics in the relativistic
  regime},}\ }\href {\doibase 10.1103/RevModPhys.78.309} {\bibfield  {journal}
  {\bibinfo  {journal} {Rev. Mod. Phys.}\ }\textbf {\bibinfo {volume} {78}},\
  \bibinfo {pages} {309--371} (\bibinfo {year} {2006})}\BibitemShut {NoStop}%
\bibitem [{\citenamefont {Pukhov}, \citenamefont {Sheng},\ and\ \citenamefont
  {Meyer-ter Vehn}(1999)}]{Pukhov_PhysPlasmas_1999}%
  \BibitemOpen
  \bibfield  {author} {\bibinfo {author} {\bibfnamefont {A.}~\bibnamefont
  {Pukhov}}, \bibinfo {author} {\bibfnamefont {Z.-M.}\ \bibnamefont {Sheng}}, \
  and\ \bibinfo {author} {\bibfnamefont {J.}~\bibnamefont {Meyer-ter Vehn}},\
  }\bibfield  {title} {\enquote {\bibinfo {title} {{Particle acceleration in
  relativistic laser channels}},}\ }\href {\doibase 10.1063/1.873242}
  {\bibfield  {journal} {\bibinfo  {journal} {Physics of Plasmas}\ }\textbf
  {\bibinfo {volume} {6}},\ \bibinfo {pages} {2847--2854} (\bibinfo {year}
  {1999})},\ \Eprint
  {http://arxiv.org/abs/https://pubs.aip.org/aip/pop/article-pdf/6/7/2847/12474740/2847\_1\_online.pdf}
  {https://pubs.aip.org/aip/pop/article-pdf/6/7/2847/12474740/2847\_1\_online.pdf}
  \BibitemShut {NoStop}%
\bibitem [{\citenamefont {Pukhov}(2002)}]{Pukhov_Review_2003}%
  \BibitemOpen
  \bibfield  {author} {\bibinfo {author} {\bibfnamefont {A.}~\bibnamefont
  {Pukhov}},\ }\bibfield  {title} {\enquote {\bibinfo {title} {Strong field
  interaction of laser radiation},}\ }\href {\doibase
  10.1088/0034-4885/66/1/202} {\bibfield  {journal} {\bibinfo  {journal}
  {Reports on Progress in Physics}\ }\textbf {\bibinfo {volume} {66}},\
  \bibinfo {pages} {47} (\bibinfo {year} {2002})}\BibitemShut {NoStop}%
\bibitem [{\citenamefont {Rosmej}\ \emph {et~al.}(2020)\citenamefont {Rosmej},
  \citenamefont {Gyrdymov}, \citenamefont {Günther}, \citenamefont {Andreev},
  \citenamefont {Tavana}, \citenamefont {Neumayer}, \citenamefont {Zähter},
  \citenamefont {Zahn}, \citenamefont {Popov}, \citenamefont {Borisenko},
  \citenamefont {Kantsyrev}, \citenamefont {Skobliakov}, \citenamefont
  {Panyushkin}, \citenamefont {Bogdanov}, \citenamefont {Consoli},
  \citenamefont {Shen},\ and\ \citenamefont {Pukhov}}]{Rosmej_PPCF_2020}%
  \BibitemOpen
  \bibfield  {author} {\bibinfo {author} {\bibfnamefont {O.~N.}\ \bibnamefont
  {Rosmej}}, \bibinfo {author} {\bibfnamefont {M.}~\bibnamefont {Gyrdymov}},
  \bibinfo {author} {\bibfnamefont {M.~M.}\ \bibnamefont {Günther}}, \bibinfo
  {author} {\bibfnamefont {N.~E.}\ \bibnamefont {Andreev}}, \bibinfo {author}
  {\bibfnamefont {P.}~\bibnamefont {Tavana}}, \bibinfo {author} {\bibfnamefont
  {P.}~\bibnamefont {Neumayer}}, \bibinfo {author} {\bibfnamefont
  {S.}~\bibnamefont {Zähter}}, \bibinfo {author} {\bibfnamefont
  {N.}~\bibnamefont {Zahn}}, \bibinfo {author} {\bibfnamefont {V.~S.}\
  \bibnamefont {Popov}}, \bibinfo {author} {\bibfnamefont {N.~G.}\ \bibnamefont
  {Borisenko}}, \bibinfo {author} {\bibfnamefont {A.}~\bibnamefont
  {Kantsyrev}}, \bibinfo {author} {\bibfnamefont {A.}~\bibnamefont
  {Skobliakov}}, \bibinfo {author} {\bibfnamefont {V.}~\bibnamefont
  {Panyushkin}}, \bibinfo {author} {\bibfnamefont {A.}~\bibnamefont
  {Bogdanov}}, \bibinfo {author} {\bibfnamefont {F.}~\bibnamefont {Consoli}},
  \bibinfo {author} {\bibfnamefont {X.~F.}\ \bibnamefont {Shen}}, \ and\
  \bibinfo {author} {\bibfnamefont {A.}~\bibnamefont {Pukhov}},\ }\bibfield
  {title} {\enquote {\bibinfo {title} {High-current laser-driven beams of
  relativistic electrons for high energy density research},}\ }\href {\doibase
  10.1088/1361-6587/abb24e} {\bibfield  {journal} {\bibinfo  {journal} {Plasma
  Physics and Controlled Fusion}\ }\textbf {\bibinfo {volume} {62}},\ \bibinfo
  {pages} {115024} (\bibinfo {year} {2020})}\BibitemShut {NoStop}%
\bibitem [{\citenamefont {Esarey}, \citenamefont {Schroeder},\ and\
  \citenamefont {Leemans}(2009)}]{Esarey_RevModPhys_2009}%
  \BibitemOpen
  \bibfield  {author} {\bibinfo {author} {\bibfnamefont {E.}~\bibnamefont
  {Esarey}}, \bibinfo {author} {\bibfnamefont {C.~B.}\ \bibnamefont
  {Schroeder}}, \ and\ \bibinfo {author} {\bibfnamefont {W.~P.}\ \bibnamefont
  {Leemans}},\ }\bibfield  {title} {\enquote {\bibinfo {title} {Physics of
  laser-driven plasma-based electron accelerators},}\ }\href {\doibase
  10.1103/RevModPhys.81.1229} {\bibfield  {journal} {\bibinfo  {journal} {Rev.
  Mod. Phys.}\ }\textbf {\bibinfo {volume} {81}},\ \bibinfo {pages}
  {1229--1285} (\bibinfo {year} {2009})}\BibitemShut {NoStop}%
\bibitem [{\citenamefont {Maksimchuk}\ \emph {et~al.}(2000)\citenamefont
  {Maksimchuk}, \citenamefont {Gu}, \citenamefont {Flippo}, \citenamefont
  {Umstadter},\ and\ \citenamefont {Bychenkov}}]{Maksimchuk_PRL_2000}%
  \BibitemOpen
  \bibfield  {author} {\bibinfo {author} {\bibfnamefont {A.}~\bibnamefont
  {Maksimchuk}}, \bibinfo {author} {\bibfnamefont {S.}~\bibnamefont {Gu}},
  \bibinfo {author} {\bibfnamefont {K.}~\bibnamefont {Flippo}}, \bibinfo
  {author} {\bibfnamefont {D.}~\bibnamefont {Umstadter}}, \ and\ \bibinfo
  {author} {\bibfnamefont {V.~Y.}\ \bibnamefont {Bychenkov}},\ }\bibfield
  {title} {{\selectlanguage {english}\enquote {\bibinfo {title} {Forward ion
  acceleration in thin films driven by a high-intensity laser},}\ }}\href
  {\doibase 10.1103/PhysRevLett.84.4108} {\bibfield  {journal} {\bibinfo
  {journal} {Phys. Rev. Lett.}\ }\textbf {\bibinfo {volume} {84}},\ \bibinfo
  {pages} {4108--4111} (\bibinfo {year} {2000})}\BibitemShut {NoStop}%
\bibitem [{\citenamefont {Snavely}\ \emph {et~al.}(2000)\citenamefont
  {Snavely}, \citenamefont {Key}, \citenamefont {Hatchett}, \citenamefont
  {Cowan}, \citenamefont {Roth}, \citenamefont {Phillips}, \citenamefont
  {Stoyer}, \citenamefont {Henry}, \citenamefont {Sangster}, \citenamefont
  {Singh}, \citenamefont {Wilks}, \citenamefont {MacKinnon}, \citenamefont
  {Offenberger}, \citenamefont {Pennington}, \citenamefont {Yasuike},
  \citenamefont {Langdon}, \citenamefont {Lasinski}, \citenamefont {Johnson},
  \citenamefont {Perry},\ and\ \citenamefont {Campbell}}]{Snavely_PRL_2000}%
  \BibitemOpen
  \bibfield  {author} {\bibinfo {author} {\bibfnamefont {R.~A.}\ \bibnamefont
  {Snavely}}, \bibinfo {author} {\bibfnamefont {M.~H.}\ \bibnamefont {Key}},
  \bibinfo {author} {\bibfnamefont {S.~P.}\ \bibnamefont {Hatchett}}, \bibinfo
  {author} {\bibfnamefont {T.~E.}\ \bibnamefont {Cowan}}, \bibinfo {author}
  {\bibfnamefont {M.}~\bibnamefont {Roth}}, \bibinfo {author} {\bibfnamefont
  {T.~W.}\ \bibnamefont {Phillips}}, \bibinfo {author} {\bibfnamefont {M.~A.}\
  \bibnamefont {Stoyer}}, \bibinfo {author} {\bibfnamefont {E.~A.}\
  \bibnamefont {Henry}}, \bibinfo {author} {\bibfnamefont {T.~C.}\ \bibnamefont
  {Sangster}}, \bibinfo {author} {\bibfnamefont {M.~S.}\ \bibnamefont {Singh}},
  \bibinfo {author} {\bibfnamefont {S.~C.}\ \bibnamefont {Wilks}}, \bibinfo
  {author} {\bibfnamefont {A.}~\bibnamefont {MacKinnon}}, \bibinfo {author}
  {\bibfnamefont {A.}~\bibnamefont {Offenberger}}, \bibinfo {author}
  {\bibfnamefont {D.~M.}\ \bibnamefont {Pennington}}, \bibinfo {author}
  {\bibfnamefont {K.}~\bibnamefont {Yasuike}}, \bibinfo {author} {\bibfnamefont
  {A.~B.}\ \bibnamefont {Langdon}}, \bibinfo {author} {\bibfnamefont {B.~F.}\
  \bibnamefont {Lasinski}}, \bibinfo {author} {\bibfnamefont {J.}~\bibnamefont
  {Johnson}}, \bibinfo {author} {\bibfnamefont {M.~D.}\ \bibnamefont {Perry}},
  \ and\ \bibinfo {author} {\bibfnamefont {E.~M.}\ \bibnamefont {Campbell}},\
  }\bibfield  {title} {\enquote {\bibinfo {title} {Intense high-energy proton
  beams from petawatt-laser irradiation of solids},}\ }\href {\doibase
  10.1103/PhysRevLett.85.2945} {\bibfield  {journal} {\bibinfo  {journal}
  {Phys. Rev. Lett.}\ }\textbf {\bibinfo {volume} {85}},\ \bibinfo {pages}
  {2945--2948} (\bibinfo {year} {2000})}\BibitemShut {NoStop}%
\bibitem [{\citenamefont {Roth}\ and\ \citenamefont
  {Schollmeier}(2016)}]{Roth_2016}%
  \BibitemOpen
  \bibfield  {author} {\bibinfo {author} {\bibfnamefont {M.}~\bibnamefont
  {Roth}}\ and\ \bibinfo {author} {\bibfnamefont {M.}~\bibnamefont
  {Schollmeier}},\ }\bibfield  {title} {{\selectlanguage {english}\enquote
  {\bibinfo {title} {Ion acceleration—target normal sheath acceleration},}\
  }}\href {\doibase 10.5170/CERN-2016-001.231} {\bibfield  {journal} {\bibinfo
  {journal} {CERN Yellow Reports}\ }\textbf {\bibinfo {volume} {1 (2016):
  Proceedings of the 2014 CAS-CERN Accelerator School: Plasma Wake
  Acceleration}} (\bibinfo {year} {2016}),\
  10.5170/CERN-2016-001.231}\BibitemShut {NoStop}%
\bibitem [{\citenamefont {Katsouleas}\ \emph {et~al.}(1997)\citenamefont
  {Katsouleas}, \citenamefont {Lee}, \citenamefont {Chattopadhyay},
  \citenamefont {Leemand}, \citenamefont {Assmann}, \citenamefont {Chen},
  \citenamefont {Decker}, \citenamefont {Iverson}, \citenamefont {Kotseroglou},
  \citenamefont {Raimondi}, \citenamefont {Raubenheimer}, \citenamefont
  {Eokni}, \citenamefont {Siemann}, \citenamefont {Walz}, \citenamefont
  {Whittum}, \citenamefont {Clayton}, \citenamefont {Joshi}, \citenamefont
  {Marsh}, \citenamefont {Mori},\ and\ \citenamefont
  {Wang}}]{Katsouleas_IEEE_1997}%
  \BibitemOpen
  \bibfield  {author} {\bibinfo {author} {\bibfnamefont {T.}~\bibnamefont
  {Katsouleas}}, \bibinfo {author} {\bibfnamefont {S.}~\bibnamefont {Lee}},
  \bibinfo {author} {\bibfnamefont {S.}~\bibnamefont {Chattopadhyay}}, \bibinfo
  {author} {\bibfnamefont {W.}~\bibnamefont {Leemand}}, \bibinfo {author}
  {\bibfnamefont {R.}~\bibnamefont {Assmann}}, \bibinfo {author} {\bibfnamefont
  {P.}~\bibnamefont {Chen}}, \bibinfo {author} {\bibfnamefont {F.}~\bibnamefont
  {Decker}}, \bibinfo {author} {\bibfnamefont {R.}~\bibnamefont {Iverson}},
  \bibinfo {author} {\bibfnamefont {T.}~\bibnamefont {Kotseroglou}}, \bibinfo
  {author} {\bibfnamefont {P.}~\bibnamefont {Raimondi}}, \bibinfo {author}
  {\bibfnamefont {T.}~\bibnamefont {Raubenheimer}}, \bibinfo {author}
  {\bibfnamefont {S.}~\bibnamefont {Eokni}}, \bibinfo {author} {\bibfnamefont
  {R.}~\bibnamefont {Siemann}}, \bibinfo {author} {\bibfnamefont
  {D.}~\bibnamefont {Walz}}, \bibinfo {author} {\bibfnamefont {D.}~\bibnamefont
  {Whittum}}, \bibinfo {author} {\bibfnamefont {C.}~\bibnamefont {Clayton}},
  \bibinfo {author} {\bibfnamefont {C.}~\bibnamefont {Joshi}}, \bibinfo
  {author} {\bibfnamefont {K.}~\bibnamefont {Marsh}}, \bibinfo {author}
  {\bibfnamefont {W.}~\bibnamefont {Mori}}, \ and\ \bibinfo {author}
  {\bibfnamefont {G.}~\bibnamefont {Wang}},\ }\bibfield  {title} {\enquote
  {\bibinfo {title} {A proposal for a 1 gev plasma-wakefield acceleration
  experiment at slac},}\ }in\ \href {\doibase 10.1109/PAC.1997.749806} {\emph
  {\bibinfo {booktitle} {Proceedings of the 1997 Particle Accelerator
  Conference (Cat. No.97CH36167)}}},\ Vol.~\bibinfo {volume} {1}\ (\bibinfo
  {year} {1997})\ pp.\ \bibinfo {pages} {687--689 vol.1}\BibitemShut {NoStop}%
\bibitem [{\citenamefont {G{\"u}nther}\ \emph {et~al.}(2022)\citenamefont
  {G{\"u}nther}, \citenamefont {Rosmej}, \citenamefont {Tavana}, \citenamefont
  {Gyrdymov}, \citenamefont {Skobliakov}, \citenamefont {Kantsyrev},
  \citenamefont {Z{\"a}hter}, \citenamefont {Borisenko}, \citenamefont
  {Pukhov},\ and\ \citenamefont {Andreev}}]{Gunther_NatCommun_2022}%
  \BibitemOpen
  \bibfield  {author} {\bibinfo {author} {\bibfnamefont {M.~M.}\ \bibnamefont
  {G{\"u}nther}}, \bibinfo {author} {\bibfnamefont {O.~N.}\ \bibnamefont
  {Rosmej}}, \bibinfo {author} {\bibfnamefont {P.}~\bibnamefont {Tavana}},
  \bibinfo {author} {\bibfnamefont {M.}~\bibnamefont {Gyrdymov}}, \bibinfo
  {author} {\bibfnamefont {A.}~\bibnamefont {Skobliakov}}, \bibinfo {author}
  {\bibfnamefont {A.}~\bibnamefont {Kantsyrev}}, \bibinfo {author}
  {\bibfnamefont {S.}~\bibnamefont {Z{\"a}hter}}, \bibinfo {author}
  {\bibfnamefont {N.~G.}\ \bibnamefont {Borisenko}}, \bibinfo {author}
  {\bibfnamefont {A.}~\bibnamefont {Pukhov}}, \ and\ \bibinfo {author}
  {\bibfnamefont {N.~E.}\ \bibnamefont {Andreev}},\ }\bibfield  {title}
  {\enquote {\bibinfo {title} {Forward-looking insights in laser-generated
  ultra-intense $\gamma$-ray and neutron sources for nuclear application and
  science},}\ }\href {\doibase 10.1038/s41467-021-27694-7} {\bibfield
  {journal} {\bibinfo  {journal} {Nature Communications}\ }\textbf {\bibinfo
  {volume} {13}},\ \bibinfo {pages} {170} (\bibinfo {year} {2022})}\BibitemShut
  {NoStop}%
\bibitem [{\citenamefont {Tavana}\ \emph {et~al.}(2023)\citenamefont {Tavana},
  \citenamefont {Bukharskii}, \citenamefont {Gyrdymov}, \citenamefont
  {Spillmann}, \citenamefont {G{\"u}nther}, \citenamefont {Cikhardt},
  \citenamefont {Z{\"a}hter}, \citenamefont {Borisenko}, \citenamefont
  {Korneev}, \citenamefont {Jacoby}, \citenamefont {Spielmann}, \citenamefont
  {Andreev},\ and\ \citenamefont {Rosmej}}]{Tavana_Frontiers_2023}%
  \BibitemOpen
  \bibfield  {author} {\bibinfo {author} {\bibfnamefont {P.}~\bibnamefont
  {Tavana}}, \bibinfo {author} {\bibfnamefont {N.}~\bibnamefont {Bukharskii}},
  \bibinfo {author} {\bibfnamefont {M.}~\bibnamefont {Gyrdymov}}, \bibinfo
  {author} {\bibfnamefont {U.}~\bibnamefont {Spillmann}}, \bibinfo {author}
  {\bibfnamefont {M.~M.}\ \bibnamefont {G{\"u}nther}}, \bibinfo {author}
  {\bibfnamefont {J.}~\bibnamefont {Cikhardt}}, \bibinfo {author}
  {\bibfnamefont {S.}~\bibnamefont {Z{\"a}hter}}, \bibinfo {author}
  {\bibfnamefont {N.~G.}\ \bibnamefont {Borisenko}}, \bibinfo {author}
  {\bibfnamefont {P.}~\bibnamefont {Korneev}}, \bibinfo {author} {\bibfnamefont
  {J.}~\bibnamefont {Jacoby}}, \bibinfo {author} {\bibfnamefont
  {C.}~\bibnamefont {Spielmann}}, \bibinfo {author} {\bibfnamefont {N.~E.}\
  \bibnamefont {Andreev}}, \ and\ \bibinfo {author} {\bibfnamefont {O.~N.}\
  \bibnamefont {Rosmej}},\ }\bibfield  {title} {\enquote {\bibinfo {title}
  {Ultra-high efficiency bremsstrahlung production in interaction of direct
  laser accelerated electrons with high-z material},}\ }\href {\doibase
  10.3389/fphy.2023.1178967} {\bibfield  {journal} {\bibinfo  {journal}
  {Frontiers in Physics}\ }\textbf {\bibinfo {volume} {11}} (\bibinfo {year}
  {2023}),\ 10.3389/fphy.2023.1178967}\BibitemShut {NoStop}%
\bibitem [{\citenamefont {Ma}\ \emph {et~al.}(2019)\citenamefont {Ma},
  \citenamefont {Lan}, \citenamefont {Liu}, \citenamefont {Wu}, \citenamefont
  {Xu}, \citenamefont {Zhu},\ and\ \citenamefont {Luo}}]{Ma_MRE_2019}%
  \BibitemOpen
  \bibfield  {author} {\bibinfo {author} {\bibfnamefont {Z.}~\bibnamefont
  {Ma}}, \bibinfo {author} {\bibfnamefont {H.}~\bibnamefont {Lan}}, \bibinfo
  {author} {\bibfnamefont {W.}~\bibnamefont {Liu}}, \bibinfo {author}
  {\bibfnamefont {S.}~\bibnamefont {Wu}}, \bibinfo {author} {\bibfnamefont
  {Y.}~\bibnamefont {Xu}}, \bibinfo {author} {\bibfnamefont {Z.}~\bibnamefont
  {Zhu}}, \ and\ \bibinfo {author} {\bibfnamefont {W.}~\bibnamefont {Luo}},\
  }\bibfield  {title} {\enquote {\bibinfo {title} {{Photonuclear production of
  medical isotopes 62,64Cu using intense laser-plasma electron source}},}\
  }\href {\doibase 10.1063/1.5100925} {\bibfield  {journal} {\bibinfo
  {journal} {Matter and Radiation at Extremes}\ }\textbf {\bibinfo {volume}
  {4}} (\bibinfo {year} {2019}),\ 10.1063/1.5100925},\ \bibinfo {note}
  {064401},\ \Eprint
  {http://arxiv.org/abs/https://pubs.aip.org/aip/mre/article-pdf/doi/10.1063/1.5100925/15691613/064401\_1\_online.pdf}
  {https://pubs.aip.org/aip/mre/article-pdf/doi/10.1063/1.5100925/15691613/064401\_1\_online.pdf}
  \BibitemShut {NoStop}%
\bibitem [{\citenamefont {Nedorezov}, \citenamefont {Rykovanov},\ and\
  \citenamefont {Savel’ev}(2021)}]{Nedorezov_2021}%
  \BibitemOpen
  \bibfield  {author} {\bibinfo {author} {\bibfnamefont {V.~G.}\ \bibnamefont
  {Nedorezov}}, \bibinfo {author} {\bibfnamefont {S.~G.}\ \bibnamefont
  {Rykovanov}}, \ and\ \bibinfo {author} {\bibfnamefont {A.~B.}\ \bibnamefont
  {Savel’ev}},\ }\bibfield  {title} {\enquote {\bibinfo {title} {Nuclear
  photonics: results and prospects},}\ }\href {\doibase
  10.3367/UFNe.2021.03.038960} {\bibfield  {journal} {\bibinfo  {journal}
  {Physics-Uspekhi}\ }\textbf {\bibinfo {volume} {64}},\ \bibinfo {pages}
  {1214} (\bibinfo {year} {2021})}\BibitemShut {NoStop}%
\bibitem [{\citenamefont {Willingale}\ \emph {et~al.}(2010)\citenamefont
  {Willingale}, \citenamefont {Nilson}, \citenamefont {Kaluza}, \citenamefont
  {Dangor}, \citenamefont {Evans}, \citenamefont {Fernandes}, \citenamefont
  {Haines}, \citenamefont {Kamperidis}, \citenamefont {Kingham}, \citenamefont
  {Ridgers}, \citenamefont {Sherlock}, \citenamefont {Thomas}, \citenamefont
  {Wei}, \citenamefont {Najmudin}, \citenamefont {Krushelnick}, \citenamefont
  {Bandyopadhyay}, \citenamefont {Notley}, \citenamefont {Minardi},
  \citenamefont {Tatarakis},\ and\ \citenamefont
  {Rozmus}}]{Willingale_PhysPlasmas_2010}%
  \BibitemOpen
  \bibfield  {author} {\bibinfo {author} {\bibfnamefont {L.}~\bibnamefont
  {Willingale}}, \bibinfo {author} {\bibfnamefont {P.~M.}\ \bibnamefont
  {Nilson}}, \bibinfo {author} {\bibfnamefont {M.~C.}\ \bibnamefont {Kaluza}},
  \bibinfo {author} {\bibfnamefont {A.~E.}\ \bibnamefont {Dangor}}, \bibinfo
  {author} {\bibfnamefont {R.~G.}\ \bibnamefont {Evans}}, \bibinfo {author}
  {\bibfnamefont {P.}~\bibnamefont {Fernandes}}, \bibinfo {author}
  {\bibfnamefont {M.~G.}\ \bibnamefont {Haines}}, \bibinfo {author}
  {\bibfnamefont {C.}~\bibnamefont {Kamperidis}}, \bibinfo {author}
  {\bibfnamefont {R.~J.}\ \bibnamefont {Kingham}}, \bibinfo {author}
  {\bibfnamefont {C.~P.}\ \bibnamefont {Ridgers}}, \bibinfo {author}
  {\bibfnamefont {M.}~\bibnamefont {Sherlock}}, \bibinfo {author}
  {\bibfnamefont {A.~G.~R.}\ \bibnamefont {Thomas}}, \bibinfo {author}
  {\bibfnamefont {M.~S.}\ \bibnamefont {Wei}}, \bibinfo {author} {\bibfnamefont
  {Z.}~\bibnamefont {Najmudin}}, \bibinfo {author} {\bibfnamefont
  {K.}~\bibnamefont {Krushelnick}}, \bibinfo {author} {\bibfnamefont
  {S.}~\bibnamefont {Bandyopadhyay}}, \bibinfo {author} {\bibfnamefont
  {M.}~\bibnamefont {Notley}}, \bibinfo {author} {\bibfnamefont
  {S.}~\bibnamefont {Minardi}}, \bibinfo {author} {\bibfnamefont
  {M.}~\bibnamefont {Tatarakis}}, \ and\ \bibinfo {author} {\bibfnamefont
  {W.}~\bibnamefont {Rozmus}},\ }\bibfield  {title} {\enquote {\bibinfo {title}
  {{Proton deflectometry of a magnetic reconnection geometry}},}\ }\href
  {\doibase 10.1063/1.3377787} {\bibfield  {journal} {\bibinfo  {journal}
  {Physics of Plasmas}\ }\textbf {\bibinfo {volume} {17}},\ \bibinfo {pages}
  {043104} (\bibinfo {year} {2010})},\ \Eprint
  {http://arxiv.org/abs/https://pubs.aip.org/aip/pop/article-pdf/doi/10.1063/1.3377787/13637568/043104\_1\_online.pdf}
  {https://pubs.aip.org/aip/pop/article-pdf/doi/10.1063/1.3377787/13637568/043104\_1\_online.pdf}
  \BibitemShut {NoStop}%
\bibitem [{\citenamefont {Law}\ \emph {et~al.}(2016{\natexlab{a}})\citenamefont
  {Law}, \citenamefont {Bailly-Grandvaux}, \citenamefont {Morace},
  \citenamefont {Sakata}, \citenamefont {Matsuo}, \citenamefont {Kojima},
  \citenamefont {Lee}, \citenamefont {Vaisseau}, \citenamefont {Arikawa},
  \citenamefont {Yogo}, \citenamefont {Kondo}, \citenamefont {Zhang},
  \citenamefont {Bellei}, \citenamefont {Santos}, \citenamefont {Fujioka},\
  and\ \citenamefont {Azechi}}]{Law_APL_2016}%
  \BibitemOpen
  \bibfield  {author} {\bibinfo {author} {\bibfnamefont {K.~F.~F.}\
  \bibnamefont {Law}}, \bibinfo {author} {\bibfnamefont {M.}~\bibnamefont
  {Bailly-Grandvaux}}, \bibinfo {author} {\bibfnamefont {A.}~\bibnamefont
  {Morace}}, \bibinfo {author} {\bibfnamefont {S.}~\bibnamefont {Sakata}},
  \bibinfo {author} {\bibfnamefont {K.}~\bibnamefont {Matsuo}}, \bibinfo
  {author} {\bibfnamefont {S.}~\bibnamefont {Kojima}}, \bibinfo {author}
  {\bibfnamefont {S.}~\bibnamefont {Lee}}, \bibinfo {author} {\bibfnamefont
  {X.}~\bibnamefont {Vaisseau}}, \bibinfo {author} {\bibfnamefont
  {Y.}~\bibnamefont {Arikawa}}, \bibinfo {author} {\bibfnamefont
  {A.}~\bibnamefont {Yogo}}, \bibinfo {author} {\bibfnamefont {K.}~\bibnamefont
  {Kondo}}, \bibinfo {author} {\bibfnamefont {Z.}~\bibnamefont {Zhang}},
  \bibinfo {author} {\bibfnamefont {C.}~\bibnamefont {Bellei}}, \bibinfo
  {author} {\bibfnamefont {J.~J.}\ \bibnamefont {Santos}}, \bibinfo {author}
  {\bibfnamefont {S.}~\bibnamefont {Fujioka}}, \ and\ \bibinfo {author}
  {\bibfnamefont {H.}~\bibnamefont {Azechi}},\ }\bibfield  {title} {\enquote
  {\bibinfo {title} {{Direct measurement of kilo-tesla level magnetic field
  generated with laser-driven capacitor-coil target by proton
  deflectometry}},}\ }\href {\doibase 10.1063/1.4943078} {\bibfield  {journal}
  {\bibinfo  {journal} {Applied Physics Letters}\ }\textbf {\bibinfo {volume}
  {108}},\ \bibinfo {pages} {091104} (\bibinfo {year} {2016}{\natexlab{a}})},\
  \Eprint
  {http://arxiv.org/abs/https://pubs.aip.org/aip/apl/article-pdf/doi/10.1063/1.4943078/13933517/091104\_1\_online.pdf}
  {https://pubs.aip.org/aip/apl/article-pdf/doi/10.1063/1.4943078/13933517/091104\_1\_online.pdf}
  \BibitemShut {NoStop}%
\bibitem [{\citenamefont {Gao}\ \emph {et~al.}(2016)\citenamefont {Gao},
  \citenamefont {Ji}, \citenamefont {Fiksel}, \citenamefont {Fox},
  \citenamefont {Evans},\ and\ \citenamefont {Alfonso}}]{Gao_PhysPlasmas_2016}%
  \BibitemOpen
  \bibfield  {author} {\bibinfo {author} {\bibfnamefont {L.}~\bibnamefont
  {Gao}}, \bibinfo {author} {\bibfnamefont {H.}~\bibnamefont {Ji}}, \bibinfo
  {author} {\bibfnamefont {G.}~\bibnamefont {Fiksel}}, \bibinfo {author}
  {\bibfnamefont {W.}~\bibnamefont {Fox}}, \bibinfo {author} {\bibfnamefont
  {M.}~\bibnamefont {Evans}}, \ and\ \bibinfo {author} {\bibfnamefont
  {N.}~\bibnamefont {Alfonso}},\ }\bibfield  {title} {\enquote {\bibinfo
  {title} {{Ultrafast proton radiography of the magnetic fields generated by a
  laser-driven coil current}},}\ }\href {\doibase 10.1063/1.4945643} {\bibfield
   {journal} {\bibinfo  {journal} {Physics of Plasmas}\ }\textbf {\bibinfo
  {volume} {23}},\ \bibinfo {pages} {043106} (\bibinfo {year} {2016})},\
  \Eprint
  {http://arxiv.org/abs/https://pubs.aip.org/aip/pop/article-pdf/doi/10.1063/1.4945643/15599806/043106\_1\_online.pdf}
  {https://pubs.aip.org/aip/pop/article-pdf/doi/10.1063/1.4945643/15599806/043106\_1\_online.pdf}
  \BibitemShut {NoStop}%
\bibitem [{\citenamefont {Palmer}\ \emph {et~al.}(2019)\citenamefont {Palmer},
  \citenamefont {Campbell}, \citenamefont {Ma}, \citenamefont {Antonelli},
  \citenamefont {Bott}, \citenamefont {Gregori}, \citenamefont {Halliday},
  \citenamefont {Katzir}, \citenamefont {Kordell}, \citenamefont {Krushelnick},
  \citenamefont {Lebedev}, \citenamefont {Montgomery}, \citenamefont {Notley},
  \citenamefont {Carroll}, \citenamefont {Ridgers}, \citenamefont
  {Schekochihin}, \citenamefont {Streeter}, \citenamefont {Thomas},
  \citenamefont {Tubman}, \citenamefont {Woolsey},\ and\ \citenamefont
  {Willingale}}]{Palmer_PhysPlasmas_2019}%
  \BibitemOpen
  \bibfield  {author} {\bibinfo {author} {\bibfnamefont {C.~A.~J.}\
  \bibnamefont {Palmer}}, \bibinfo {author} {\bibfnamefont {P.~T.}\
  \bibnamefont {Campbell}}, \bibinfo {author} {\bibfnamefont {Y.}~\bibnamefont
  {Ma}}, \bibinfo {author} {\bibfnamefont {L.}~\bibnamefont {Antonelli}},
  \bibinfo {author} {\bibfnamefont {A.~F.~A.}\ \bibnamefont {Bott}}, \bibinfo
  {author} {\bibfnamefont {G.}~\bibnamefont {Gregori}}, \bibinfo {author}
  {\bibfnamefont {J.}~\bibnamefont {Halliday}}, \bibinfo {author}
  {\bibfnamefont {Y.}~\bibnamefont {Katzir}}, \bibinfo {author} {\bibfnamefont
  {P.}~\bibnamefont {Kordell}}, \bibinfo {author} {\bibfnamefont
  {K.}~\bibnamefont {Krushelnick}}, \bibinfo {author} {\bibfnamefont {S.~V.}\
  \bibnamefont {Lebedev}}, \bibinfo {author} {\bibfnamefont {E.}~\bibnamefont
  {Montgomery}}, \bibinfo {author} {\bibfnamefont {M.}~\bibnamefont {Notley}},
  \bibinfo {author} {\bibfnamefont {D.~C.}\ \bibnamefont {Carroll}}, \bibinfo
  {author} {\bibfnamefont {C.~P.}\ \bibnamefont {Ridgers}}, \bibinfo {author}
  {\bibfnamefont {A.~A.}\ \bibnamefont {Schekochihin}}, \bibinfo {author}
  {\bibfnamefont {M.~J.~V.}\ \bibnamefont {Streeter}}, \bibinfo {author}
  {\bibfnamefont {A.~G.~R.}\ \bibnamefont {Thomas}}, \bibinfo {author}
  {\bibfnamefont {E.~R.}\ \bibnamefont {Tubman}}, \bibinfo {author}
  {\bibfnamefont {N.}~\bibnamefont {Woolsey}}, \ and\ \bibinfo {author}
  {\bibfnamefont {L.}~\bibnamefont {Willingale}},\ }\bibfield  {title}
  {\enquote {\bibinfo {title} {{Field reconstruction from proton radiography of
  intense laser driven magnetic reconnection}},}\ }\href {\doibase
  10.1063/1.5092733} {\bibfield  {journal} {\bibinfo  {journal} {Physics of
  Plasmas}\ }\textbf {\bibinfo {volume} {26}},\ \bibinfo {pages} {083109}
  (\bibinfo {year} {2019})},\ \Eprint
  {http://arxiv.org/abs/https://pubs.aip.org/aip/pop/article-pdf/doi/10.1063/1.5092733/15856787/083109\_1\_online.pdf}
  {https://pubs.aip.org/aip/pop/article-pdf/doi/10.1063/1.5092733/15856787/083109\_1\_online.pdf}
  \BibitemShut {NoStop}%
\bibitem [{\citenamefont {Bradford}\ \emph {et~al.}(2020)\citenamefont
  {Bradford}, \citenamefont {Read}, \citenamefont {Ehret}, \citenamefont
  {Antonelli}, \citenamefont {Khan}, \citenamefont {Booth}, \citenamefont
  {Glize}, \citenamefont {Carroll}, \citenamefont {Clarke}, \citenamefont
  {Heathcote},\ and\ \citenamefont {et~al.}}]{Bradford_HPLSE_2020}%
  \BibitemOpen
  \bibfield  {author} {\bibinfo {author} {\bibfnamefont {P.}~\bibnamefont
  {Bradford}}, \bibinfo {author} {\bibfnamefont {M.~P.}\ \bibnamefont {Read}},
  \bibinfo {author} {\bibfnamefont {M.}~\bibnamefont {Ehret}}, \bibinfo
  {author} {\bibfnamefont {L.}~\bibnamefont {Antonelli}}, \bibinfo {author}
  {\bibfnamefont {M.}~\bibnamefont {Khan}}, \bibinfo {author} {\bibfnamefont
  {N.}~\bibnamefont {Booth}}, \bibinfo {author} {\bibfnamefont
  {K.}~\bibnamefont {Glize}}, \bibinfo {author} {\bibfnamefont
  {D.}~\bibnamefont {Carroll}}, \bibinfo {author} {\bibfnamefont {R.~J.}\
  \bibnamefont {Clarke}}, \bibinfo {author} {\bibfnamefont {R.}~\bibnamefont
  {Heathcote}}, \ and\ \bibinfo {author} {\bibnamefont {et~al.}},\ }\bibfield
  {title} {\enquote {\bibinfo {title} {Proton deflectometry of a capacitor coil
  target along two axes},}\ }\href {\doibase 10.1017/hpl.2020.9} {\bibfield
  {journal} {\bibinfo  {journal} {High Power Laser Science and Engineering}\
  }\textbf {\bibinfo {volume} {8}},\ \bibinfo {pages} {e11} (\bibinfo {year}
  {2020})}\BibitemShut {NoStop}%
\bibitem [{\citenamefont {Patel}\ \emph {et~al.}(2003)\citenamefont {Patel},
  \citenamefont {Mackinnon}, \citenamefont {Key}, \citenamefont {Cowan},
  \citenamefont {Foord}, \citenamefont {Allen}, \citenamefont {Price},
  \citenamefont {Ruhl}, \citenamefont {Springer},\ and\ \citenamefont
  {Stephens}}]{Patel_PhysRevLett_2003}%
  \BibitemOpen
  \bibfield  {author} {\bibinfo {author} {\bibfnamefont {P.~K.}\ \bibnamefont
  {Patel}}, \bibinfo {author} {\bibfnamefont {A.~J.}\ \bibnamefont
  {Mackinnon}}, \bibinfo {author} {\bibfnamefont {M.~H.}\ \bibnamefont {Key}},
  \bibinfo {author} {\bibfnamefont {T.~E.}\ \bibnamefont {Cowan}}, \bibinfo
  {author} {\bibfnamefont {M.~E.}\ \bibnamefont {Foord}}, \bibinfo {author}
  {\bibfnamefont {M.}~\bibnamefont {Allen}}, \bibinfo {author} {\bibfnamefont
  {D.~F.}\ \bibnamefont {Price}}, \bibinfo {author} {\bibfnamefont
  {H.}~\bibnamefont {Ruhl}}, \bibinfo {author} {\bibfnamefont {P.~T.}\
  \bibnamefont {Springer}}, \ and\ \bibinfo {author} {\bibfnamefont
  {R.}~\bibnamefont {Stephens}},\ }\bibfield  {title} {\enquote {\bibinfo
  {title} {Isochoric heating of solid-density matter with an ultrafast proton
  beam},}\ }\href {\doibase 10.1103/PhysRevLett.91.125004} {\bibfield
  {journal} {\bibinfo  {journal} {Phys. Rev. Lett.}\ }\textbf {\bibinfo
  {volume} {91}},\ \bibinfo {pages} {125004} (\bibinfo {year}
  {2003})}\BibitemShut {NoStop}%
\bibitem [{\citenamefont {Dyer}\ \emph {et~al.}(2008)\citenamefont {Dyer},
  \citenamefont {Bernstein}, \citenamefont {Cho}, \citenamefont {Osterholz},
  \citenamefont {Grigsby}, \citenamefont {Dalton}, \citenamefont {Shepherd},
  \citenamefont {Ping}, \citenamefont {Chen}, \citenamefont {Widmann},\ and\
  \citenamefont {Ditmire}}]{Dyer_PhysRevLett_2008}%
  \BibitemOpen
  \bibfield  {author} {\bibinfo {author} {\bibfnamefont {G.~M.}\ \bibnamefont
  {Dyer}}, \bibinfo {author} {\bibfnamefont {A.~C.}\ \bibnamefont {Bernstein}},
  \bibinfo {author} {\bibfnamefont {B.~I.}\ \bibnamefont {Cho}}, \bibinfo
  {author} {\bibfnamefont {J.}~\bibnamefont {Osterholz}}, \bibinfo {author}
  {\bibfnamefont {W.}~\bibnamefont {Grigsby}}, \bibinfo {author} {\bibfnamefont
  {A.}~\bibnamefont {Dalton}}, \bibinfo {author} {\bibfnamefont
  {R.}~\bibnamefont {Shepherd}}, \bibinfo {author} {\bibfnamefont
  {Y.}~\bibnamefont {Ping}}, \bibinfo {author} {\bibfnamefont {H.}~\bibnamefont
  {Chen}}, \bibinfo {author} {\bibfnamefont {K.}~\bibnamefont {Widmann}}, \
  and\ \bibinfo {author} {\bibfnamefont {T.}~\bibnamefont {Ditmire}},\
  }\bibfield  {title} {\enquote {\bibinfo {title} {Equation-of-state
  measurement of dense plasmas heated with fast protons},}\ }\href {\doibase
  10.1103/PhysRevLett.101.015002} {\bibfield  {journal} {\bibinfo  {journal}
  {Phys. Rev. Lett.}\ }\textbf {\bibinfo {volume} {101}},\ \bibinfo {pages}
  {015002} (\bibinfo {year} {2008})}\BibitemShut {NoStop}%
\bibitem [{\citenamefont {Temporal}, \citenamefont {Honrubia},\ and\
  \citenamefont {Atzeni}(2002)}]{Temporal_PhysPlasmas_2002}%
  \BibitemOpen
  \bibfield  {author} {\bibinfo {author} {\bibfnamefont {M.}~\bibnamefont
  {Temporal}}, \bibinfo {author} {\bibfnamefont {J.~J.}\ \bibnamefont
  {Honrubia}}, \ and\ \bibinfo {author} {\bibfnamefont {S.}~\bibnamefont
  {Atzeni}},\ }\bibfield  {title} {\enquote {\bibinfo {title} {{Numerical study
  of fast ignition of ablatively imploded deuterium–tritium fusion capsules
  by ultra-intense proton beams}},}\ }\href {\doibase 10.1063/1.1482375}
  {\bibfield  {journal} {\bibinfo  {journal} {Physics of Plasmas}\ }\textbf
  {\bibinfo {volume} {9}},\ \bibinfo {pages} {3098--3107} (\bibinfo {year}
  {2002})},\ \Eprint
  {http://arxiv.org/abs/https://pubs.aip.org/aip/pop/article-pdf/9/7/3098/19222104/3098\_1\_online.pdf}
  {https://pubs.aip.org/aip/pop/article-pdf/9/7/3098/19222104/3098\_1\_online.pdf}
  \BibitemShut {NoStop}%
\bibitem [{\citenamefont {Bychenkov}\ \emph {et~al.}(2001)\citenamefont
  {Bychenkov}, \citenamefont {Rozmus}, \citenamefont {Maksimchuk},
  \citenamefont {Umstadter},\ and\ \citenamefont
  {Capjack}}]{Bychenkov_PlasmaPhysRep_2001}%
  \BibitemOpen
  \bibfield  {author} {\bibinfo {author} {\bibfnamefont {V.~Y.}\ \bibnamefont
  {Bychenkov}}, \bibinfo {author} {\bibfnamefont {W.}~\bibnamefont {Rozmus}},
  \bibinfo {author} {\bibfnamefont {A.}~\bibnamefont {Maksimchuk}}, \bibinfo
  {author} {\bibfnamefont {D.}~\bibnamefont {Umstadter}}, \ and\ \bibinfo
  {author} {\bibfnamefont {C.~E.}\ \bibnamefont {Capjack}},\ }\bibfield
  {title} {\enquote {\bibinfo {title} {Fast ignitor concept with light ions},}\
  }\href {\doibase 10.1134/1.1426135} {\bibfield  {journal} {\bibinfo
  {journal} {Plasma Physics Reports}\ }\textbf {\bibinfo {volume} {27}},\
  \bibinfo {pages} {1017--1020} (\bibinfo {year} {2001})}\BibitemShut {NoStop}%
\bibitem [{\citenamefont {Bulanov}\ \emph {et~al.}(2002)\citenamefont
  {Bulanov}, \citenamefont {Esirkepov}, \citenamefont {Khoroshkov},
  \citenamefont {Kuznetsov},\ and\ \citenamefont
  {Pegoraro}}]{Bulanov_PhysLettA_2002}%
  \BibitemOpen
  \bibfield  {author} {\bibinfo {author} {\bibfnamefont {S.}~\bibnamefont
  {Bulanov}}, \bibinfo {author} {\bibfnamefont {T.}~\bibnamefont {Esirkepov}},
  \bibinfo {author} {\bibfnamefont {V.}~\bibnamefont {Khoroshkov}}, \bibinfo
  {author} {\bibfnamefont {A.}~\bibnamefont {Kuznetsov}}, \ and\ \bibinfo
  {author} {\bibfnamefont {F.}~\bibnamefont {Pegoraro}},\ }\bibfield  {title}
  {\enquote {\bibinfo {title} {Oncological hadrontherapy with laser ion
  accelerators},}\ }\href {\doibase
  https://doi.org/10.1016/S0375-9601(02)00521-2} {\bibfield  {journal}
  {\bibinfo  {journal} {Physics Letters A}\ }\textbf {\bibinfo {volume}
  {299}},\ \bibinfo {pages} {240--247} (\bibinfo {year} {2002})}\BibitemShut
  {NoStop}%
\bibitem [{\citenamefont {Fourkal}\ \emph {et~al.}(2002)\citenamefont
  {Fourkal}, \citenamefont {Shahine}, \citenamefont {Ding}, \citenamefont {Li},
  \citenamefont {Tajima},\ and\ \citenamefont {Ma}}]{Fourkal_MedPhys_2002}%
  \BibitemOpen
  \bibfield  {author} {\bibinfo {author} {\bibfnamefont {E.}~\bibnamefont
  {Fourkal}}, \bibinfo {author} {\bibfnamefont {B.}~\bibnamefont {Shahine}},
  \bibinfo {author} {\bibfnamefont {M.}~\bibnamefont {Ding}}, \bibinfo {author}
  {\bibfnamefont {J.~S.}\ \bibnamefont {Li}}, \bibinfo {author} {\bibfnamefont
  {T.}~\bibnamefont {Tajima}}, \ and\ \bibinfo {author} {\bibfnamefont {C.-M.}\
  \bibnamefont {Ma}},\ }\bibfield  {title} {\enquote {\bibinfo {title}
  {Particle in cell simulation of laser-accelerated proton beams for radiation
  therapy},}\ }\href {\doibase https://doi.org/10.1118/1.1521122} {\bibfield
  {journal} {\bibinfo  {journal} {Medical Physics}\ }\textbf {\bibinfo {volume}
  {29}},\ \bibinfo {pages} {2788--2798} (\bibinfo {year} {2002})},\ \Eprint
  {http://arxiv.org/abs/https://aapm.onlinelibrary.wiley.com/doi/pdf/10.1118/1.1521122}
  {https://aapm.onlinelibrary.wiley.com/doi/pdf/10.1118/1.1521122} \BibitemShut
  {NoStop}%
\bibitem [{\citenamefont {Fourkal}\ \emph {et~al.}(2003)\citenamefont
  {Fourkal}, \citenamefont {Li}, \citenamefont {Ding}, \citenamefont {Tajima},\
  and\ \citenamefont {Ma}}]{Fourkal_MedPhys_2003}%
  \BibitemOpen
  \bibfield  {author} {\bibinfo {author} {\bibfnamefont {E.}~\bibnamefont
  {Fourkal}}, \bibinfo {author} {\bibfnamefont {J.~S.}\ \bibnamefont {Li}},
  \bibinfo {author} {\bibfnamefont {M.}~\bibnamefont {Ding}}, \bibinfo {author}
  {\bibfnamefont {T.}~\bibnamefont {Tajima}}, \ and\ \bibinfo {author}
  {\bibfnamefont {C.-M.}\ \bibnamefont {Ma}},\ }\bibfield  {title} {\enquote
  {\bibinfo {title} {Particle selection for laser-accelerated proton therapy
  feasibility study},}\ }\href {\doibase https://doi.org/10.1118/1.1586268}
  {\bibfield  {journal} {\bibinfo  {journal} {Medical Physics}\ }\textbf
  {\bibinfo {volume} {30}},\ \bibinfo {pages} {1660--1670} (\bibinfo {year}
  {2003})},\ \Eprint
  {http://arxiv.org/abs/https://aapm.onlinelibrary.wiley.com/doi/pdf/10.1118/1.1586268}
  {https://aapm.onlinelibrary.wiley.com/doi/pdf/10.1118/1.1586268} \BibitemShut
  {NoStop}%
\bibitem [{\citenamefont {Bailly-Grandvaux}\ \emph {et~al.}(2018)\citenamefont
  {Bailly-Grandvaux}, \citenamefont {Santos}, \citenamefont {Bellei},
  \citenamefont {Forestier-Colleoni}, \citenamefont {Fujioka}, \citenamefont
  {Giuffrida}, \citenamefont {Honrubia}, \citenamefont {Batani}, \citenamefont
  {Bouillaud}, \citenamefont {Chevrot}, \citenamefont {Cross}, \citenamefont
  {Crowston}, \citenamefont {Dorard}, \citenamefont {Dubois}, \citenamefont
  {Ehret}, \citenamefont {Gregori}, \citenamefont {Hulin}, \citenamefont
  {Kojima}, \citenamefont {Loyez}, \citenamefont {Marqu{\`e}s}, \citenamefont
  {Morace}, \citenamefont {Nicola{\"i}}, \citenamefont {Roth}, \citenamefont
  {Sakata}, \citenamefont {Schaumann}, \citenamefont {Serres}, \citenamefont
  {Servel}, \citenamefont {Tikhonchuk}, \citenamefont {Woolsey},\ and\
  \citenamefont {Zhang}}]{Bailly-Grandvaux_NatCommun_2018}%
  \BibitemOpen
  \bibfield  {author} {\bibinfo {author} {\bibfnamefont {M.}~\bibnamefont
  {Bailly-Grandvaux}}, \bibinfo {author} {\bibfnamefont {J.~J.}\ \bibnamefont
  {Santos}}, \bibinfo {author} {\bibfnamefont {C.}~\bibnamefont {Bellei}},
  \bibinfo {author} {\bibfnamefont {P.}~\bibnamefont {Forestier-Colleoni}},
  \bibinfo {author} {\bibfnamefont {S.}~\bibnamefont {Fujioka}}, \bibinfo
  {author} {\bibfnamefont {L.}~\bibnamefont {Giuffrida}}, \bibinfo {author}
  {\bibfnamefont {J.~J.}\ \bibnamefont {Honrubia}}, \bibinfo {author}
  {\bibfnamefont {D.}~\bibnamefont {Batani}}, \bibinfo {author} {\bibfnamefont
  {R.}~\bibnamefont {Bouillaud}}, \bibinfo {author} {\bibfnamefont
  {M.}~\bibnamefont {Chevrot}}, \bibinfo {author} {\bibfnamefont {J.~E.}\
  \bibnamefont {Cross}}, \bibinfo {author} {\bibfnamefont {R.}~\bibnamefont
  {Crowston}}, \bibinfo {author} {\bibfnamefont {S.}~\bibnamefont {Dorard}},
  \bibinfo {author} {\bibfnamefont {J.-L.}\ \bibnamefont {Dubois}}, \bibinfo
  {author} {\bibfnamefont {M.}~\bibnamefont {Ehret}}, \bibinfo {author}
  {\bibfnamefont {G.}~\bibnamefont {Gregori}}, \bibinfo {author} {\bibfnamefont
  {S.}~\bibnamefont {Hulin}}, \bibinfo {author} {\bibfnamefont
  {S.}~\bibnamefont {Kojima}}, \bibinfo {author} {\bibfnamefont
  {E.}~\bibnamefont {Loyez}}, \bibinfo {author} {\bibfnamefont {J.-R.}\
  \bibnamefont {Marqu{\`e}s}}, \bibinfo {author} {\bibfnamefont
  {A.}~\bibnamefont {Morace}}, \bibinfo {author} {\bibfnamefont
  {P.}~\bibnamefont {Nicola{\"i}}}, \bibinfo {author} {\bibfnamefont
  {M.}~\bibnamefont {Roth}}, \bibinfo {author} {\bibfnamefont {S.}~\bibnamefont
  {Sakata}}, \bibinfo {author} {\bibfnamefont {G.}~\bibnamefont {Schaumann}},
  \bibinfo {author} {\bibfnamefont {F.}~\bibnamefont {Serres}}, \bibinfo
  {author} {\bibfnamefont {J.}~\bibnamefont {Servel}}, \bibinfo {author}
  {\bibfnamefont {V.~T.}\ \bibnamefont {Tikhonchuk}}, \bibinfo {author}
  {\bibfnamefont {N.}~\bibnamefont {Woolsey}}, \ and\ \bibinfo {author}
  {\bibfnamefont {Z.}~\bibnamefont {Zhang}},\ }\bibfield  {title} {\enquote
  {\bibinfo {title} {Guiding of relativistic electron beams in dense matter by
  laser-driven magnetostatic fields},}\ }\href {\doibase
  10.1038/s41467-017-02641-7} {\bibfield  {journal} {\bibinfo  {journal}
  {Nature Communications}\ }\textbf {\bibinfo {volume} {9}},\ \bibinfo {pages}
  {102} (\bibinfo {year} {2018})}\BibitemShut {NoStop}%
\bibitem [{\citenamefont {Santos}\ \emph {et~al.}(2018)\citenamefont {Santos},
  \citenamefont {Bailly-Grandvaux}, \citenamefont {Ehret}, \citenamefont
  {Arefiev}, \citenamefont {Batani}, \citenamefont {Beg}, \citenamefont
  {Calisti}, \citenamefont {Ferri}, \citenamefont {Florido}, \citenamefont
  {Forestier-Colleoni}, \citenamefont {Fujioka}, \citenamefont {Gigosos},
  \citenamefont {Giuffrida}, \citenamefont {Gremillet}, \citenamefont
  {Honrubia}, \citenamefont {Kojima}, \citenamefont {Korneev}, \citenamefont
  {Law}, \citenamefont {Marquès}, \citenamefont {Morace}, \citenamefont
  {Mossé}, \citenamefont {Peyrusse}, \citenamefont {Rose}, \citenamefont
  {Roth}, \citenamefont {Sakata}, \citenamefont {Schaumann}, \citenamefont
  {Suzuki-Vidal}, \citenamefont {Tikhonchuk}, \citenamefont {Toncian},
  \citenamefont {Woolsey},\ and\ \citenamefont
  {Zhang}}]{Santos_PhysPlasmas_2018}%
  \BibitemOpen
  \bibfield  {author} {\bibinfo {author} {\bibfnamefont {J.~J.}\ \bibnamefont
  {Santos}}, \bibinfo {author} {\bibfnamefont {M.}~\bibnamefont
  {Bailly-Grandvaux}}, \bibinfo {author} {\bibfnamefont {M.}~\bibnamefont
  {Ehret}}, \bibinfo {author} {\bibfnamefont {A.~V.}\ \bibnamefont {Arefiev}},
  \bibinfo {author} {\bibfnamefont {D.}~\bibnamefont {Batani}}, \bibinfo
  {author} {\bibfnamefont {F.~N.}\ \bibnamefont {Beg}}, \bibinfo {author}
  {\bibfnamefont {A.}~\bibnamefont {Calisti}}, \bibinfo {author} {\bibfnamefont
  {S.}~\bibnamefont {Ferri}}, \bibinfo {author} {\bibfnamefont
  {R.}~\bibnamefont {Florido}}, \bibinfo {author} {\bibfnamefont
  {P.}~\bibnamefont {Forestier-Colleoni}}, \bibinfo {author} {\bibfnamefont
  {S.}~\bibnamefont {Fujioka}}, \bibinfo {author} {\bibfnamefont {M.~A.}\
  \bibnamefont {Gigosos}}, \bibinfo {author} {\bibfnamefont {L.}~\bibnamefont
  {Giuffrida}}, \bibinfo {author} {\bibfnamefont {L.}~\bibnamefont
  {Gremillet}}, \bibinfo {author} {\bibfnamefont {J.~J.}\ \bibnamefont
  {Honrubia}}, \bibinfo {author} {\bibfnamefont {S.}~\bibnamefont {Kojima}},
  \bibinfo {author} {\bibfnamefont {P.}~\bibnamefont {Korneev}}, \bibinfo
  {author} {\bibfnamefont {K.~F.~F.}\ \bibnamefont {Law}}, \bibinfo {author}
  {\bibfnamefont {J.-R.}\ \bibnamefont {Marquès}}, \bibinfo {author}
  {\bibfnamefont {A.}~\bibnamefont {Morace}}, \bibinfo {author} {\bibfnamefont
  {C.}~\bibnamefont {Mossé}}, \bibinfo {author} {\bibfnamefont
  {O.}~\bibnamefont {Peyrusse}}, \bibinfo {author} {\bibfnamefont
  {S.}~\bibnamefont {Rose}}, \bibinfo {author} {\bibfnamefont {M.}~\bibnamefont
  {Roth}}, \bibinfo {author} {\bibfnamefont {S.}~\bibnamefont {Sakata}},
  \bibinfo {author} {\bibfnamefont {G.}~\bibnamefont {Schaumann}}, \bibinfo
  {author} {\bibfnamefont {F.}~\bibnamefont {Suzuki-Vidal}}, \bibinfo {author}
  {\bibfnamefont {V.~T.}\ \bibnamefont {Tikhonchuk}}, \bibinfo {author}
  {\bibfnamefont {T.}~\bibnamefont {Toncian}}, \bibinfo {author} {\bibfnamefont
  {N.}~\bibnamefont {Woolsey}}, \ and\ \bibinfo {author} {\bibfnamefont
  {Z.}~\bibnamefont {Zhang}},\ }\bibfield  {title} {\enquote {\bibinfo {title}
  {{Laser-driven strong magnetostatic fields with applications to charged beam
  transport and magnetized high energy-density physics}},}\ }\href {\doibase
  10.1063/1.5018735} {\bibfield  {journal} {\bibinfo  {journal} {Physics of
  Plasmas}\ }\textbf {\bibinfo {volume} {25}},\ \bibinfo {pages} {056705}
  (\bibinfo {year} {2018})},\ \Eprint
  {http://arxiv.org/abs/https://pubs.aip.org/aip/pop/article-pdf/doi/10.1063/1.5018735/14700567/056705\_1\_online.pdf}
  {https://pubs.aip.org/aip/pop/article-pdf/doi/10.1063/1.5018735/14700567/056705\_1\_online.pdf}
  \BibitemShut {NoStop}%
\bibitem [{\citenamefont {Kar}\ \emph {et~al.}(2016)\citenamefont {Kar},
  \citenamefont {Ahmed}, \citenamefont {Prasad}, \citenamefont {Cerchez},
  \citenamefont {Brauckmann}, \citenamefont {Aurand}, \citenamefont {Cantono},
  \citenamefont {Hadjisolomou}, \citenamefont {Lewis}, \citenamefont {Macchi},
  \citenamefont {Nersisyan}, \citenamefont {Robinson}, \citenamefont {Schroer},
  \citenamefont {Swantusch}, \citenamefont {Zepf}, \citenamefont {Willi},\ and\
  \citenamefont {Borghesi}}]{Kar_NatCommun_2016}%
  \BibitemOpen
  \bibfield  {author} {\bibinfo {author} {\bibfnamefont {S.}~\bibnamefont
  {Kar}}, \bibinfo {author} {\bibfnamefont {H.}~\bibnamefont {Ahmed}}, \bibinfo
  {author} {\bibfnamefont {R.}~\bibnamefont {Prasad}}, \bibinfo {author}
  {\bibfnamefont {M.}~\bibnamefont {Cerchez}}, \bibinfo {author} {\bibfnamefont
  {S.}~\bibnamefont {Brauckmann}}, \bibinfo {author} {\bibfnamefont
  {B.}~\bibnamefont {Aurand}}, \bibinfo {author} {\bibfnamefont
  {G.}~\bibnamefont {Cantono}}, \bibinfo {author} {\bibfnamefont
  {P.}~\bibnamefont {Hadjisolomou}}, \bibinfo {author} {\bibfnamefont
  {C.~L.~S.}\ \bibnamefont {Lewis}}, \bibinfo {author} {\bibfnamefont
  {A.}~\bibnamefont {Macchi}}, \bibinfo {author} {\bibfnamefont
  {G.}~\bibnamefont {Nersisyan}}, \bibinfo {author} {\bibfnamefont {A.~P.~L.}\
  \bibnamefont {Robinson}}, \bibinfo {author} {\bibfnamefont {A.~M.}\
  \bibnamefont {Schroer}}, \bibinfo {author} {\bibfnamefont {M.}~\bibnamefont
  {Swantusch}}, \bibinfo {author} {\bibfnamefont {M.}~\bibnamefont {Zepf}},
  \bibinfo {author} {\bibfnamefont {O.}~\bibnamefont {Willi}}, \ and\ \bibinfo
  {author} {\bibfnamefont {M.}~\bibnamefont {Borghesi}},\ }\bibfield  {title}
  {\enquote {\bibinfo {title} {Guided post-acceleration of laser-driven ions by
  a miniature modular structure},}\ }\href {\doibase 10.1038/ncomms10792}
  {\bibfield  {journal} {\bibinfo  {journal} {Nature Communications}\ }\textbf
  {\bibinfo {volume} {7}},\ \bibinfo {pages} {10792} (\bibinfo {year}
  {2016})}\BibitemShut {NoStop}%
\bibitem [{\citenamefont {Ahmed}\ \emph {et~al.}(2017)\citenamefont {Ahmed},
  \citenamefont {Kar}, \citenamefont {Cantono}, \citenamefont {Hadjisolomou},
  \citenamefont {Poye}, \citenamefont {Gwynne}, \citenamefont {Lewis},
  \citenamefont {Macchi}, \citenamefont {Naughton}, \citenamefont {Nersisyan},
  \citenamefont {Tikhonchuk}, \citenamefont {Willi},\ and\ \citenamefont
  {Borghesi}}]{Ahmed_SciRep_2017}%
  \BibitemOpen
  \bibfield  {author} {\bibinfo {author} {\bibfnamefont {H.}~\bibnamefont
  {Ahmed}}, \bibinfo {author} {\bibfnamefont {S.}~\bibnamefont {Kar}}, \bibinfo
  {author} {\bibfnamefont {G.}~\bibnamefont {Cantono}}, \bibinfo {author}
  {\bibfnamefont {P.}~\bibnamefont {Hadjisolomou}}, \bibinfo {author}
  {\bibfnamefont {A.}~\bibnamefont {Poye}}, \bibinfo {author} {\bibfnamefont
  {D.}~\bibnamefont {Gwynne}}, \bibinfo {author} {\bibfnamefont {C.~L.~S.}\
  \bibnamefont {Lewis}}, \bibinfo {author} {\bibfnamefont {A.}~\bibnamefont
  {Macchi}}, \bibinfo {author} {\bibfnamefont {K.}~\bibnamefont {Naughton}},
  \bibinfo {author} {\bibfnamefont {G.}~\bibnamefont {Nersisyan}}, \bibinfo
  {author} {\bibfnamefont {V.}~\bibnamefont {Tikhonchuk}}, \bibinfo {author}
  {\bibfnamefont {O.}~\bibnamefont {Willi}}, \ and\ \bibinfo {author}
  {\bibfnamefont {M.}~\bibnamefont {Borghesi}},\ }\bibfield  {title} {\enquote
  {\bibinfo {title} {Efficient post-acceleration of protons in helical coil
  targets driven by sub-ps laser pulses},}\ }\href {\doibase
  10.1038/s41598-017-06985-4} {\bibfield  {journal} {\bibinfo  {journal}
  {Scientific Reports}\ }\textbf {\bibinfo {volume} {7}},\ \bibinfo {pages}
  {10891} (\bibinfo {year} {2017})}\BibitemShut {NoStop}%
\bibitem [{\citenamefont {Liu}\ \emph {et~al.}(2024)\citenamefont {Liu},
  \citenamefont {Gao}, \citenamefont {Wu}, \citenamefont {Pan}, \citenamefont
  {Liang}, \citenamefont {Song}, \citenamefont {Xu}, \citenamefont {Shou},
  \citenamefont {Zhang}, \citenamefont {Chen}, \citenamefont {Han},
  \citenamefont {Hua}, \citenamefont {Chen}, \citenamefont {Xu}, \citenamefont
  {Mei}, \citenamefont {Wang}, \citenamefont {Peng}, \citenamefont {Zhao},
  \citenamefont {Chen}, \citenamefont {Zhao}, \citenamefont {Yan},\ and\
  \citenamefont {Ma}}]{Liu_PhysPlasmas_2024}%
  \BibitemOpen
  \bibfield  {author} {\bibinfo {author} {\bibfnamefont {Z.}~\bibnamefont
  {Liu}}, \bibinfo {author} {\bibfnamefont {Y.}~\bibnamefont {Gao}}, \bibinfo
  {author} {\bibfnamefont {Q.}~\bibnamefont {Wu}}, \bibinfo {author}
  {\bibfnamefont {Z.}~\bibnamefont {Pan}}, \bibinfo {author} {\bibfnamefont
  {Y.}~\bibnamefont {Liang}}, \bibinfo {author} {\bibfnamefont
  {T.}~\bibnamefont {Song}}, \bibinfo {author} {\bibfnamefont {T.}~\bibnamefont
  {Xu}}, \bibinfo {author} {\bibfnamefont {Y.}~\bibnamefont {Shou}}, \bibinfo
  {author} {\bibfnamefont {Y.}~\bibnamefont {Zhang}}, \bibinfo {author}
  {\bibfnamefont {H.}~\bibnamefont {Chen}}, \bibinfo {author} {\bibfnamefont
  {Q.}~\bibnamefont {Han}}, \bibinfo {author} {\bibfnamefont {C.}~\bibnamefont
  {Hua}}, \bibinfo {author} {\bibfnamefont {X.}~\bibnamefont {Chen}}, \bibinfo
  {author} {\bibfnamefont {S.}~\bibnamefont {Xu}}, \bibinfo {author}
  {\bibfnamefont {Z.}~\bibnamefont {Mei}}, \bibinfo {author} {\bibfnamefont
  {P.}~\bibnamefont {Wang}}, \bibinfo {author} {\bibfnamefont {Z.}~\bibnamefont
  {Peng}}, \bibinfo {author} {\bibfnamefont {J.}~\bibnamefont {Zhao}}, \bibinfo
  {author} {\bibfnamefont {S.}~\bibnamefont {Chen}}, \bibinfo {author}
  {\bibfnamefont {Y.}~\bibnamefont {Zhao}}, \bibinfo {author} {\bibfnamefont
  {X.}~\bibnamefont {Yan}}, \ and\ \bibinfo {author} {\bibfnamefont
  {W.}~\bibnamefont {Ma}},\ }\bibfield  {title} {\enquote {\bibinfo {title}
  {{Synergistic enhancement of laser-proton acceleration with integrated
  targets}},}\ }\href {\doibase 10.1063/5.0195634} {\bibfield  {journal}
  {\bibinfo  {journal} {Physics of Plasmas}\ }\textbf {\bibinfo {volume}
  {31}},\ \bibinfo {pages} {053106} (\bibinfo {year} {2024})},\ \Eprint
  {http://arxiv.org/abs/https://pubs.aip.org/aip/pop/article-pdf/doi/10.1063/5.0195634/19960308/053106\_1\_5.0195634.pdf}
  {https://pubs.aip.org/aip/pop/article-pdf/doi/10.1063/5.0195634/19960308/053106\_1\_5.0195634.pdf}
  \BibitemShut {NoStop}%
\bibitem [{\citenamefont {Santos}\ \emph {et~al.}(2015)\citenamefont {Santos},
  \citenamefont {{Bailly-Grandvaux}}, \citenamefont {Giuffrida}, \citenamefont
  {{Forestier-Colleoni}}, \citenamefont {Fujioka}, \citenamefont {Zhang},
  \citenamefont {Korneev}, \citenamefont {Bouillaud}, \citenamefont {Dorard},
  \citenamefont {Batani}, \citenamefont {Chevrot}, \citenamefont {Cross},
  \citenamefont {Crowston}, \citenamefont {Dubois}, \citenamefont {Gazave},
  \citenamefont {Gregori}, \citenamefont {D'humières}, \citenamefont {Hulin},
  \citenamefont {Ishihara}, \citenamefont {Kojima}, \citenamefont {Loyez},
  \citenamefont {Marquès}, \citenamefont {Morace}, \citenamefont {Nicolaï},
  \citenamefont {Peyrusse}, \citenamefont {Poyé}, \citenamefont {Raffestin},
  \citenamefont {Ribolzi}, \citenamefont {Roth}, \citenamefont {Schaumann},
  \citenamefont {Serres}, \citenamefont {Tikhonchuk}, \citenamefont {Vacar},\
  and\ \citenamefont
  {Woolsey}}]{Santos.etal_LaserdrivenPlatformGeneration_NJP-2015}%
  \BibitemOpen
  \bibfield  {author} {\bibinfo {author} {\bibfnamefont {J.~J.}\ \bibnamefont
  {Santos}}, \bibinfo {author} {\bibfnamefont {M.}~\bibnamefont
  {{Bailly-Grandvaux}}}, \bibinfo {author} {\bibfnamefont {L.}~\bibnamefont
  {Giuffrida}}, \bibinfo {author} {\bibfnamefont {P.}~\bibnamefont
  {{Forestier-Colleoni}}}, \bibinfo {author} {\bibfnamefont {S.}~\bibnamefont
  {Fujioka}}, \bibinfo {author} {\bibfnamefont {Z.}~\bibnamefont {Zhang}},
  \bibinfo {author} {\bibfnamefont {P.}~\bibnamefont {Korneev}}, \bibinfo
  {author} {\bibfnamefont {R.}~\bibnamefont {Bouillaud}}, \bibinfo {author}
  {\bibfnamefont {S.}~\bibnamefont {Dorard}}, \bibinfo {author} {\bibfnamefont
  {D.}~\bibnamefont {Batani}}, \bibinfo {author} {\bibfnamefont
  {M.}~\bibnamefont {Chevrot}}, \bibinfo {author} {\bibfnamefont {J.~E.}\
  \bibnamefont {Cross}}, \bibinfo {author} {\bibfnamefont {R.}~\bibnamefont
  {Crowston}}, \bibinfo {author} {\bibfnamefont {J.~L.}\ \bibnamefont
  {Dubois}}, \bibinfo {author} {\bibfnamefont {J.}~\bibnamefont {Gazave}},
  \bibinfo {author} {\bibfnamefont {G.}~\bibnamefont {Gregori}}, \bibinfo
  {author} {\bibfnamefont {E.}~\bibnamefont {D'humières}}, \bibinfo {author}
  {\bibfnamefont {S.}~\bibnamefont {Hulin}}, \bibinfo {author} {\bibfnamefont
  {K.}~\bibnamefont {Ishihara}}, \bibinfo {author} {\bibfnamefont
  {S.}~\bibnamefont {Kojima}}, \bibinfo {author} {\bibfnamefont
  {E.}~\bibnamefont {Loyez}}, \bibinfo {author} {\bibfnamefont {J.~R.}\
  \bibnamefont {Marquès}}, \bibinfo {author} {\bibfnamefont {A.}~\bibnamefont
  {Morace}}, \bibinfo {author} {\bibfnamefont {P.}~\bibnamefont {Nicolaï}},
  \bibinfo {author} {\bibfnamefont {O.}~\bibnamefont {Peyrusse}}, \bibinfo
  {author} {\bibfnamefont {A.}~\bibnamefont {Poyé}}, \bibinfo {author}
  {\bibfnamefont {D.}~\bibnamefont {Raffestin}}, \bibinfo {author}
  {\bibfnamefont {J.}~\bibnamefont {Ribolzi}}, \bibinfo {author} {\bibfnamefont
  {M.}~\bibnamefont {Roth}}, \bibinfo {author} {\bibfnamefont {G.}~\bibnamefont
  {Schaumann}}, \bibinfo {author} {\bibfnamefont {F.}~\bibnamefont {Serres}},
  \bibinfo {author} {\bibfnamefont {V.~T.}\ \bibnamefont {Tikhonchuk}},
  \bibinfo {author} {\bibfnamefont {P.}~\bibnamefont {Vacar}}, \ and\ \bibinfo
  {author} {\bibfnamefont {N.}~\bibnamefont {Woolsey}},\ }\bibfield  {title}
  {\enquote {\bibinfo {title} {Laser-driven platform for generation and
  characterization of strong quasi-static magnetic fields},}\ }\href {\doibase
  10.1088/1367-2630/17/8/083051} {\bibfield  {journal} {\bibinfo  {journal}
  {New Journal of Physics}\ }\textbf {\bibinfo {volume} {17}},\ \bibinfo
  {pages} {083051} (\bibinfo {year} {2015})},\ \Eprint
  {http://arxiv.org/abs/1503.00247} {arXiv:1503.00247} \BibitemShut {NoStop}%
\bibitem [{\citenamefont {Law}\ \emph {et~al.}(2016{\natexlab{b}})\citenamefont
  {Law}, \citenamefont {{Bailly-Grandvaux}}, \citenamefont {Morace},
  \citenamefont {Sakata}, \citenamefont {Matsuo}, \citenamefont {Kojima},
  \citenamefont {Lee}, \citenamefont {Vaisseau}, \citenamefont {Arikawa},
  \citenamefont {Yogo}, \citenamefont {Kondo}, \citenamefont {Zhang},
  \citenamefont {Bellei}, \citenamefont {Santos}, \citenamefont {Fujioka},\
  and\ \citenamefont {Azechi}}]{Law-apl16}%
  \BibitemOpen
  \bibfield  {author} {\bibinfo {author} {\bibfnamefont {K.~F.~F.}\
  \bibnamefont {Law}}, \bibinfo {author} {\bibfnamefont {M.}~\bibnamefont
  {{Bailly-Grandvaux}}}, \bibinfo {author} {\bibfnamefont {A.}~\bibnamefont
  {Morace}}, \bibinfo {author} {\bibfnamefont {S.}~\bibnamefont {Sakata}},
  \bibinfo {author} {\bibfnamefont {K.}~\bibnamefont {Matsuo}}, \bibinfo
  {author} {\bibfnamefont {S.}~\bibnamefont {Kojima}}, \bibinfo {author}
  {\bibfnamefont {S.}~\bibnamefont {Lee}}, \bibinfo {author} {\bibfnamefont
  {X.}~\bibnamefont {Vaisseau}}, \bibinfo {author} {\bibfnamefont
  {Y.}~\bibnamefont {Arikawa}}, \bibinfo {author} {\bibfnamefont
  {A.}~\bibnamefont {Yogo}}, \bibinfo {author} {\bibfnamefont {K.}~\bibnamefont
  {Kondo}}, \bibinfo {author} {\bibfnamefont {Z.}~\bibnamefont {Zhang}},
  \bibinfo {author} {\bibfnamefont {C.}~\bibnamefont {Bellei}}, \bibinfo
  {author} {\bibfnamefont {J.~J.}\ \bibnamefont {Santos}}, \bibinfo {author}
  {\bibfnamefont {S.}~\bibnamefont {Fujioka}}, \ and\ \bibinfo {author}
  {\bibfnamefont {H.}~\bibnamefont {Azechi}},\ }\bibfield  {title} {\enquote
  {\bibinfo {title} {Direct measurement of kilo-tesla level magnetic field
  generated with laser-driven capacitor-coil target by proton deflectometry},}\
  }\href {\doibase 10.1063/1.4943078} {\bibfield  {journal} {\bibinfo
  {journal} {Applied Physics Letters}\ }\textbf {\bibinfo {volume} {108}},\
  \bibinfo {pages} {091104} (\bibinfo {year} {2016}{\natexlab{b}})}\BibitemShut
  {NoStop}%
\bibitem [{\citenamefont {Daido}\ \emph {et~al.}(1986)\citenamefont {Daido},
  \citenamefont {Miki}, \citenamefont {Mima}, \citenamefont {Fujita},
  \citenamefont {Sawai}, \citenamefont {Fujita}, \citenamefont {Kitagawa},
  \citenamefont {Nakai},\ and\ \citenamefont {Yamanaka}}]{Daido1986}%
  \BibitemOpen
  \bibfield  {author} {\bibinfo {author} {\bibfnamefont {H.}~\bibnamefont
  {Daido}}, \bibinfo {author} {\bibfnamefont {F.}~\bibnamefont {Miki}},
  \bibinfo {author} {\bibfnamefont {K.}~\bibnamefont {Mima}}, \bibinfo {author}
  {\bibfnamefont {M.}~\bibnamefont {Fujita}}, \bibinfo {author} {\bibfnamefont
  {K.}~\bibnamefont {Sawai}}, \bibinfo {author} {\bibfnamefont
  {H.}~\bibnamefont {Fujita}}, \bibinfo {author} {\bibfnamefont
  {Y.}~\bibnamefont {Kitagawa}}, \bibinfo {author} {\bibfnamefont
  {S.}~\bibnamefont {Nakai}}, \ and\ \bibinfo {author} {\bibfnamefont
  {C.}~\bibnamefont {Yamanaka}},\ }\bibfield  {title} {\enquote {\bibinfo
  {title} {Generation of a strong magnetic field by an intense {{CO2}} laser
  pulse},}\ }\href {\doibase 10.1103/PhysRevLett.56.846} {\bibfield  {journal}
  {\bibinfo  {journal} {Physical Review Letters}\ }\textbf {\bibinfo {volume}
  {56}},\ \bibinfo {pages} {846--849} (\bibinfo {year} {1986})}\BibitemShut
  {NoStop}%
\bibitem [{\citenamefont {Courtois}\ \emph {et~al.}(2005)\citenamefont
  {Courtois}, \citenamefont {Ash}, \citenamefont {Chambers}, \citenamefont
  {Grundy},\ and\ \citenamefont {Woolsey}}]{Courtois-jap05}%
  \BibitemOpen
  \bibfield  {author} {\bibinfo {author} {\bibfnamefont {C.}~\bibnamefont
  {Courtois}}, \bibinfo {author} {\bibfnamefont {A.~D.}\ \bibnamefont {Ash}},
  \bibinfo {author} {\bibfnamefont {D.~M.}\ \bibnamefont {Chambers}}, \bibinfo
  {author} {\bibfnamefont {R.~A.~D.}\ \bibnamefont {Grundy}}, \ and\ \bibinfo
  {author} {\bibfnamefont {N.~C.}\ \bibnamefont {Woolsey}},\ }\bibfield
  {title} {\enquote {\bibinfo {title} {Creation of a uniform high
  magnetic-field strength environment for laser-driven experiments},}\ }\href
  {\doibase 10.1063/1.2035896} {\bibfield  {journal} {\bibinfo  {journal}
  {Journal of Applied Physics}\ }\textbf {\bibinfo {volume} {98}},\ \bibinfo
  {pages} {54913} (\bibinfo {year} {2005})}\BibitemShut {NoStop}%
\bibitem [{\citenamefont {Ehret}\ \emph {et~al.}(2023)\citenamefont {Ehret},
  \citenamefont {{Bailly-Grandvaux}}, \citenamefont {Korneev}, \citenamefont
  {Apiñaniz}, \citenamefont {Brabetz}, \citenamefont {Morace}, \citenamefont
  {Bradford}, \citenamefont {{d'Humières}}, \citenamefont {Schaumann},
  \citenamefont {Bagnoud}, \citenamefont {Malko}, \citenamefont {Matveevskii},
  \citenamefont {Roth}, \citenamefont {Volpe}, \citenamefont {Woolsey},\ and\
  \citenamefont {Santos}}]{Ehret.etal_GuidedElectromagneticDischarge_PoP-2023}%
  \BibitemOpen
  \bibfield  {author} {\bibinfo {author} {\bibfnamefont {M.}~\bibnamefont
  {Ehret}}, \bibinfo {author} {\bibfnamefont {M.}~\bibnamefont
  {{Bailly-Grandvaux}}}, \bibinfo {author} {\bibfnamefont {{\relax
  Ph}.}~\bibnamefont {Korneev}}, \bibinfo {author} {\bibfnamefont {J.~I.}\
  \bibnamefont {Apiñaniz}}, \bibinfo {author} {\bibfnamefont {C.}~\bibnamefont
  {Brabetz}}, \bibinfo {author} {\bibfnamefont {A.}~\bibnamefont {Morace}},
  \bibinfo {author} {\bibfnamefont {P.}~\bibnamefont {Bradford}}, \bibinfo
  {author} {\bibfnamefont {E.}~\bibnamefont {{d'Humières}}}, \bibinfo {author}
  {\bibfnamefont {G.}~\bibnamefont {Schaumann}}, \bibinfo {author}
  {\bibfnamefont {V.}~\bibnamefont {Bagnoud}}, \bibinfo {author} {\bibfnamefont
  {S.}~\bibnamefont {Malko}}, \bibinfo {author} {\bibfnamefont
  {K.}~\bibnamefont {Matveevskii}}, \bibinfo {author} {\bibfnamefont
  {M.}~\bibnamefont {Roth}}, \bibinfo {author} {\bibfnamefont {L.}~\bibnamefont
  {Volpe}}, \bibinfo {author} {\bibfnamefont {N.~C.}\ \bibnamefont {Woolsey}},
  \ and\ \bibinfo {author} {\bibfnamefont {J.~J.}\ \bibnamefont {Santos}},\
  }\bibfield  {title} {\enquote {\bibinfo {title} {Guided electromagnetic
  discharge pulses driven by short intense laser pulses: {{Characterization}}
  and modeling},}\ }\href {\doibase 10.1063/5.0124011} {\bibfield  {journal}
  {\bibinfo  {journal} {Physics of Plasmas}\ }\textbf {\bibinfo {volume}
  {30}},\ \bibinfo {pages} {013105} (\bibinfo {year} {2023})}\BibitemShut
  {NoStop}%
\bibitem [{\citenamefont {Ehret}\ \emph {et~al.}(2022)\citenamefont {Ehret},
  \citenamefont {Kochetkov}, \citenamefont {Abe}, \citenamefont {Law},
  \citenamefont {Bukharskii}, \citenamefont {Stepanischev}, \citenamefont
  {Fujioka}, \citenamefont {d'Humi\`eres}, \citenamefont {Zielbauer},
  \citenamefont {Bagnoud}, \citenamefont {Schaumann}, \citenamefont {Somekawa},
  \citenamefont {Roth}, \citenamefont {Tikhonchuk}, \citenamefont {Santos},\
  and\ \citenamefont {Korneev}}]{Ehret_PhysRevE_2022}%
  \BibitemOpen
  \bibfield  {author} {\bibinfo {author} {\bibfnamefont {M.}~\bibnamefont
  {Ehret}}, \bibinfo {author} {\bibfnamefont {Y.}~\bibnamefont {Kochetkov}},
  \bibinfo {author} {\bibfnamefont {Y.}~\bibnamefont {Abe}}, \bibinfo {author}
  {\bibfnamefont {K.~F.~F.}\ \bibnamefont {Law}}, \bibinfo {author}
  {\bibfnamefont {N.}~\bibnamefont {Bukharskii}}, \bibinfo {author}
  {\bibfnamefont {V.}~\bibnamefont {Stepanischev}}, \bibinfo {author}
  {\bibfnamefont {S.}~\bibnamefont {Fujioka}}, \bibinfo {author} {\bibfnamefont
  {E.}~\bibnamefont {d'Humi\`eres}}, \bibinfo {author} {\bibfnamefont
  {B.}~\bibnamefont {Zielbauer}}, \bibinfo {author} {\bibfnamefont
  {V.}~\bibnamefont {Bagnoud}}, \bibinfo {author} {\bibfnamefont
  {G.}~\bibnamefont {Schaumann}}, \bibinfo {author} {\bibfnamefont
  {T.}~\bibnamefont {Somekawa}}, \bibinfo {author} {\bibfnamefont
  {M.}~\bibnamefont {Roth}}, \bibinfo {author} {\bibfnamefont {V.}~\bibnamefont
  {Tikhonchuk}}, \bibinfo {author} {\bibfnamefont {J.~J.}\ \bibnamefont
  {Santos}}, \ and\ \bibinfo {author} {\bibfnamefont {P.}~\bibnamefont
  {Korneev}},\ }\bibfield  {title} {\enquote {\bibinfo {title} {Kilotesla
  plasmoid formation by a trapped relativistic laser beam},}\ }\href {\doibase
  10.1103/PhysRevE.106.045211} {\bibfield  {journal} {\bibinfo  {journal}
  {Phys. Rev. E}\ }\textbf {\bibinfo {volume} {106}},\ \bibinfo {pages}
  {045211} (\bibinfo {year} {2022})}\BibitemShut {NoStop}%
\bibitem [{\citenamefont {Bukharskii}\ and\ \citenamefont
  {Korneev}(2023)}]{Bukharskii_BullLebedevPhysInst_2023}%
  \BibitemOpen
  \bibfield  {author} {\bibinfo {author} {\bibfnamefont {N.~D.}\ \bibnamefont
  {Bukharskii}}\ and\ \bibinfo {author} {\bibfnamefont {P.~A.}\ \bibnamefont
  {Korneev}},\ }\bibfield  {title} {\enquote {\bibinfo {title} {Study of a
  highly magnetized relativistic plasma in the context of laboratory
  astrophysics and particle flow control},}\ }\href {\doibase
  10.3103/S1068335623200022} {\bibfield  {journal} {\bibinfo  {journal}
  {Bulletin of the Lebedev Physics Institute}\ }\textbf {\bibinfo {volume}
  {50}},\ \bibinfo {pages} {S869--S877} (\bibinfo {year} {2023})}\BibitemShut
  {NoStop}%
\bibitem [{\citenamefont {Korneev}, \citenamefont {D'Humières},\ and\
  \citenamefont
  {Tikhonchuk}(2015)}]{Korneev.etal_GigagaussscaleQuasistaticMagnetic_PRE-2015}%
  \BibitemOpen
  \bibfield  {author} {\bibinfo {author} {\bibfnamefont {{\relax
  Ph}.}~\bibnamefont {Korneev}}, \bibinfo {author} {\bibfnamefont
  {E.}~\bibnamefont {D'Humières}}, \ and\ \bibinfo {author} {\bibfnamefont
  {V.}~\bibnamefont {Tikhonchuk}},\ }\bibfield  {title} {\enquote {\bibinfo
  {title} {Gigagauss-scale quasistatic magnetic field generation in a
  snail-shaped target},}\ }\href {\doibase 10.1103/PhysRevE.91.043107}
  {\bibfield  {journal} {\bibinfo  {journal} {Physical Review E}\ }\textbf
  {\bibinfo {volume} {91}},\ \bibinfo {pages} {43107} (\bibinfo {year}
  {2015})},\ \Eprint {http://arxiv.org/abs/1410.0053} {arXiv:1410.0053}
  \BibitemShut {NoStop}%
\bibitem [{\citenamefont
  {Korneev}(2017)}]{Korneev_MagnetizedPlasmaStructures_JPCS-2017}%
  \BibitemOpen
  \bibfield  {author} {\bibinfo {author} {\bibfnamefont {P.}~\bibnamefont
  {Korneev}},\ }\bibfield  {title} {\enquote {\bibinfo {title} {Magnetized
  plasma structures in laser-irradiated curved targets},}\ }\href {\doibase
  10.1088/1742-6596/788/1/012042} {\bibfield  {journal} {\bibinfo  {journal}
  {Journal of Physics: Conference Series}\ }\textbf {\bibinfo {volume} {788}},\
  \bibinfo {pages} {012042} (\bibinfo {year} {2017})}\BibitemShut {NoStop}%
\bibitem [{\citenamefont {Van~Rossum}\ and\ \citenamefont
  {Drake}(2009)}]{Python}%
  \BibitemOpen
  \bibfield  {author} {\bibinfo {author} {\bibfnamefont {G.}~\bibnamefont
  {Van~Rossum}}\ and\ \bibinfo {author} {\bibfnamefont {F.~L.}\ \bibnamefont
  {Drake}},\ }\href@noop {} {{\selectlanguage {english}\emph {\bibinfo {title}
  {Python 3 Reference Manual}}}}\ (\bibinfo  {publisher} {CreateSpace},\
  \bibinfo {address} {Scotts Valley, CA},\ \bibinfo {year} {2009})\BibitemShut
  {NoStop}%
\bibitem [{\citenamefont {Harris}\ \emph {et~al.}(2020)\citenamefont {Harris},
  \citenamefont {Millman}, \citenamefont {van~der Walt}, \citenamefont
  {Gommers}, \citenamefont {Virtanen}, \citenamefont {Cournapeau},
  \citenamefont {Wieser}, \citenamefont {Taylor}, \citenamefont {Berg},
  \citenamefont {Smith}, \citenamefont {Kern}, \citenamefont {Picus},
  \citenamefont {Hoyer}, \citenamefont {van Kerkwijk}, \citenamefont {Brett},
  \citenamefont {Haldane}, \citenamefont {del R{\'{i}}o}, \citenamefont
  {Wiebe}, \citenamefont {Peterson}, \citenamefont {G{\'{e}}rard-Marchant},
  \citenamefont {Sheppard}, \citenamefont {Reddy}, \citenamefont {Weckesser},
  \citenamefont {Abbasi}, \citenamefont {Gohlke},\ and\ \citenamefont
  {Oliphant}}]{Numpy}%
  \BibitemOpen
  \bibfield  {author} {\bibinfo {author} {\bibfnamefont {C.~R.}\ \bibnamefont
  {Harris}}, \bibinfo {author} {\bibfnamefont {K.~J.}\ \bibnamefont {Millman}},
  \bibinfo {author} {\bibfnamefont {S.~J.}\ \bibnamefont {van~der Walt}},
  \bibinfo {author} {\bibfnamefont {R.}~\bibnamefont {Gommers}}, \bibinfo
  {author} {\bibfnamefont {P.}~\bibnamefont {Virtanen}}, \bibinfo {author}
  {\bibfnamefont {D.}~\bibnamefont {Cournapeau}}, \bibinfo {author}
  {\bibfnamefont {E.}~\bibnamefont {Wieser}}, \bibinfo {author} {\bibfnamefont
  {J.}~\bibnamefont {Taylor}}, \bibinfo {author} {\bibfnamefont
  {S.}~\bibnamefont {Berg}}, \bibinfo {author} {\bibfnamefont {N.~J.}\
  \bibnamefont {Smith}}, \bibinfo {author} {\bibfnamefont {R.}~\bibnamefont
  {Kern}}, \bibinfo {author} {\bibfnamefont {M.}~\bibnamefont {Picus}},
  \bibinfo {author} {\bibfnamefont {S.}~\bibnamefont {Hoyer}}, \bibinfo
  {author} {\bibfnamefont {M.~H.}\ \bibnamefont {van Kerkwijk}}, \bibinfo
  {author} {\bibfnamefont {M.}~\bibnamefont {Brett}}, \bibinfo {author}
  {\bibfnamefont {A.}~\bibnamefont {Haldane}}, \bibinfo {author} {\bibfnamefont
  {J.~F.}\ \bibnamefont {del R{\'{i}}o}}, \bibinfo {author} {\bibfnamefont
  {M.}~\bibnamefont {Wiebe}}, \bibinfo {author} {\bibfnamefont
  {P.}~\bibnamefont {Peterson}}, \bibinfo {author} {\bibfnamefont
  {P.}~\bibnamefont {G{\'{e}}rard-Marchant}}, \bibinfo {author} {\bibfnamefont
  {K.}~\bibnamefont {Sheppard}}, \bibinfo {author} {\bibfnamefont
  {T.}~\bibnamefont {Reddy}}, \bibinfo {author} {\bibfnamefont
  {W.}~\bibnamefont {Weckesser}}, \bibinfo {author} {\bibfnamefont
  {H.}~\bibnamefont {Abbasi}}, \bibinfo {author} {\bibfnamefont
  {C.}~\bibnamefont {Gohlke}}, \ and\ \bibinfo {author} {\bibfnamefont {T.~E.}\
  \bibnamefont {Oliphant}},\ }\bibfield  {title} {{\selectlanguage
  {english}\enquote {\bibinfo {title} {Array programming with {NumPy}},}\
  }}\href {\doibase 10.1038/s41586-020-2649-2} {\bibfield  {journal} {\bibinfo
  {journal} {Nature}\ }\textbf {\bibinfo {volume} {585}},\ \bibinfo {pages}
  {357--362} (\bibinfo {year} {2020})}\BibitemShut {NoStop}%
\bibitem [{\citenamefont {Lam}, \citenamefont {Pitrou},\ and\ \citenamefont
  {Seibert}(2015)}]{Numba}%
  \BibitemOpen
  \bibfield  {author} {\bibinfo {author} {\bibfnamefont {S.~K.}\ \bibnamefont
  {Lam}}, \bibinfo {author} {\bibfnamefont {A.}~\bibnamefont {Pitrou}}, \ and\
  \bibinfo {author} {\bibfnamefont {S.}~\bibnamefont {Seibert}},\ }\bibfield
  {title} {{\selectlanguage {english}\enquote {\bibinfo {title} {Numba: a
  llvm-based python jit compiler},}\ }}in\ \href {\doibase
  10.1145/2833157.2833162} {{\selectlanguage {english}\emph {\bibinfo
  {booktitle} {Proceedings of the Second Workshop on the LLVM Compiler
  Infrastructure in HPC}}}},\ \bibinfo {series and number} {LLVM '15}\
  (\bibinfo  {publisher} {Association for Computing Machinery},\ \bibinfo
  {address} {New York, NY, USA},\ \bibinfo {year} {2015})\BibitemShut {NoStop}%
\bibitem [{\citenamefont {Boris}(1970)}]{Boris_1970}%
  \BibitemOpen
  \bibfield  {author} {\bibinfo {author} {\bibfnamefont {J.~P.}\ \bibnamefont
  {Boris}},\ }\bibfield  {title} {\enquote {\bibinfo {title} {Relativistic
  plasma simulation-optimization of a hybrid code},}\ }\href@noop {} {\bibfield
   {journal} {\bibinfo  {journal} {Proceeding of Fourth Conference on Numerical
  Simulations of Plasmas}\ } (\bibinfo {year} {1970})}\BibitemShut {NoStop}%
\bibitem [{\citenamefont {Reiser}(2008)}]{Reiser}%
  \BibitemOpen
  \bibfield  {author} {\bibinfo {author} {\bibfnamefont {M.}~\bibnamefont
  {Reiser}},\ }\href {http://dx.doi.org/10.1002/9783527622047} {\enquote
  {\bibinfo {title} {Theory and design of charged particle beams},}\ }
  (\bibinfo {year} {2008})\BibitemShut {NoStop}%
\bibitem [{\citenamefont {Borghesi}\ \emph {et~al.}(2004)\citenamefont
  {Borghesi}, \citenamefont {Mackinnon}, \citenamefont {Campbell},
  \citenamefont {Hicks}, \citenamefont {Kar}, \citenamefont {Patel},
  \citenamefont {Price}, \citenamefont {Romagnani}, \citenamefont {Schiavi},\
  and\ \citenamefont {Willi}}]{Borghesi_PRL_2004}%
  \BibitemOpen
  \bibfield  {author} {\bibinfo {author} {\bibfnamefont {M.}~\bibnamefont
  {Borghesi}}, \bibinfo {author} {\bibfnamefont {A.~J.}\ \bibnamefont
  {Mackinnon}}, \bibinfo {author} {\bibfnamefont {D.~H.}\ \bibnamefont
  {Campbell}}, \bibinfo {author} {\bibfnamefont {D.~G.}\ \bibnamefont {Hicks}},
  \bibinfo {author} {\bibfnamefont {S.}~\bibnamefont {Kar}}, \bibinfo {author}
  {\bibfnamefont {P.~K.}\ \bibnamefont {Patel}}, \bibinfo {author}
  {\bibfnamefont {D.}~\bibnamefont {Price}}, \bibinfo {author} {\bibfnamefont
  {L.}~\bibnamefont {Romagnani}}, \bibinfo {author} {\bibfnamefont
  {A.}~\bibnamefont {Schiavi}}, \ and\ \bibinfo {author} {\bibfnamefont
  {O.}~\bibnamefont {Willi}},\ }\bibfield  {title} {\enquote {\bibinfo {title}
  {Multi-mev proton source investigations in ultraintense laser-foil
  interactions},}\ }\href {\doibase 10.1103/PhysRevLett.92.055003} {\bibfield
  {journal} {\bibinfo  {journal} {Phys. Rev. Lett.}\ }\textbf {\bibinfo
  {volume} {92}},\ \bibinfo {pages} {055003} (\bibinfo {year}
  {2004})}\BibitemShut {NoStop}%
\bibitem [{\citenamefont {Kugland}\ \emph {et~al.}(2012)\citenamefont
  {Kugland}, \citenamefont {Ryutov}, \citenamefont {Plechaty}, \citenamefont
  {Ross},\ and\ \citenamefont {Park}}]{Kugland_RSI_2012}%
  \BibitemOpen
  \bibfield  {author} {\bibinfo {author} {\bibfnamefont {N.~L.}\ \bibnamefont
  {Kugland}}, \bibinfo {author} {\bibfnamefont {D.~D.}\ \bibnamefont {Ryutov}},
  \bibinfo {author} {\bibfnamefont {C.}~\bibnamefont {Plechaty}}, \bibinfo
  {author} {\bibfnamefont {J.~S.}\ \bibnamefont {Ross}}, \ and\ \bibinfo
  {author} {\bibfnamefont {H.-S.}\ \bibnamefont {Park}},\ }\bibfield  {title}
  {\enquote {\bibinfo {title} {{Invited Article: Relation between electric and
  magnetic field structures and their proton-beam images}},}\ }\href {\doibase
  10.1063/1.4750234} {\bibfield  {journal} {\bibinfo  {journal} {Review of
  Scientific Instruments}\ }\textbf {\bibinfo {volume} {83}},\ \bibinfo {pages}
  {101301} (\bibinfo {year} {2012})},\ \Eprint
  {http://arxiv.org/abs/https://pubs.aip.org/aip/rsi/article-pdf/doi/10.1063/1.4750234/14133841/101301\_1\_online.pdf}
  {https://pubs.aip.org/aip/rsi/article-pdf/doi/10.1063/1.4750234/14133841/101301\_1\_online.pdf}
  \BibitemShut {NoStop}%
\bibitem [{\citenamefont {Hegelich}\ \emph {et~al.}(2006)\citenamefont
  {Hegelich}, \citenamefont {Albright}, \citenamefont {Cobble}, \citenamefont
  {Flippo}, \citenamefont {Letzring}, \citenamefont {Paffett}, \citenamefont
  {Ruhl}, \citenamefont {Schreiber}, \citenamefont {Schulze},\ and\
  \citenamefont {Fern{\'a}ndez}}]{Hegelich_Nature_2006}%
  \BibitemOpen
  \bibfield  {author} {\bibinfo {author} {\bibfnamefont {B.~M.}\ \bibnamefont
  {Hegelich}}, \bibinfo {author} {\bibfnamefont {B.~J.}\ \bibnamefont
  {Albright}}, \bibinfo {author} {\bibfnamefont {J.}~\bibnamefont {Cobble}},
  \bibinfo {author} {\bibfnamefont {K.}~\bibnamefont {Flippo}}, \bibinfo
  {author} {\bibfnamefont {S.}~\bibnamefont {Letzring}}, \bibinfo {author}
  {\bibfnamefont {M.}~\bibnamefont {Paffett}}, \bibinfo {author} {\bibfnamefont
  {H.}~\bibnamefont {Ruhl}}, \bibinfo {author} {\bibfnamefont {J.}~\bibnamefont
  {Schreiber}}, \bibinfo {author} {\bibfnamefont {R.~K.}\ \bibnamefont
  {Schulze}}, \ and\ \bibinfo {author} {\bibfnamefont {J.~C.}\ \bibnamefont
  {Fern{\'a}ndez}},\ }\bibfield  {title} {\enquote {\bibinfo {title} {Laser
  acceleration of quasi-monoenergetic mev ion beams},}\ }\href {\doibase
  10.1038/nature04400} {\bibfield  {journal} {\bibinfo  {journal} {Nature}\
  }\textbf {\bibinfo {volume} {439}},\ \bibinfo {pages} {441--444} (\bibinfo
  {year} {2006})}\BibitemShut {NoStop}%
\bibitem [{\citenamefont {Schwoerer}\ \emph {et~al.}(2006)\citenamefont
  {Schwoerer}, \citenamefont {Pfotenhauer}, \citenamefont {J{\"a}ckel},
  \citenamefont {Amthor}, \citenamefont {Liesfeld}, \citenamefont {Ziegler},
  \citenamefont {Sauerbrey}, \citenamefont {Ledingham},\ and\ \citenamefont
  {Esirkepov}}]{Schwoerer_Nature_2006}%
  \BibitemOpen
  \bibfield  {author} {\bibinfo {author} {\bibfnamefont {H.}~\bibnamefont
  {Schwoerer}}, \bibinfo {author} {\bibfnamefont {S.}~\bibnamefont
  {Pfotenhauer}}, \bibinfo {author} {\bibfnamefont {O.}~\bibnamefont
  {J{\"a}ckel}}, \bibinfo {author} {\bibfnamefont {K.-U.}\ \bibnamefont
  {Amthor}}, \bibinfo {author} {\bibfnamefont {B.}~\bibnamefont {Liesfeld}},
  \bibinfo {author} {\bibfnamefont {W.}~\bibnamefont {Ziegler}}, \bibinfo
  {author} {\bibfnamefont {R.}~\bibnamefont {Sauerbrey}}, \bibinfo {author}
  {\bibfnamefont {K.~W.~D.}\ \bibnamefont {Ledingham}}, \ and\ \bibinfo
  {author} {\bibfnamefont {T.}~\bibnamefont {Esirkepov}},\ }\bibfield  {title}
  {\enquote {\bibinfo {title} {Laser-plasma acceleration of quasi-monoenergetic
  protons from microstructured targets},}\ }\href {\doibase
  10.1038/nature04492} {\bibfield  {journal} {\bibinfo  {journal} {Nature}\
  }\textbf {\bibinfo {volume} {439}},\ \bibinfo {pages} {445--448} (\bibinfo
  {year} {2006})}\BibitemShut {NoStop}%
\bibitem [{\citenamefont {Ter-Avetisyan}\ \emph {et~al.}(2006)\citenamefont
  {Ter-Avetisyan}, \citenamefont {Schn\"urer}, \citenamefont {Nickles},
  \citenamefont {Kalashnikov}, \citenamefont {Risse}, \citenamefont {Sokollik},
  \citenamefont {Sandner}, \citenamefont {Andreev},\ and\ \citenamefont
  {Tikhonchuk}}]{TerAvetisyan_PRL_2006}%
  \BibitemOpen
  \bibfield  {author} {\bibinfo {author} {\bibfnamefont {S.}~\bibnamefont
  {Ter-Avetisyan}}, \bibinfo {author} {\bibfnamefont {M.}~\bibnamefont
  {Schn\"urer}}, \bibinfo {author} {\bibfnamefont {P.~V.}\ \bibnamefont
  {Nickles}}, \bibinfo {author} {\bibfnamefont {M.}~\bibnamefont
  {Kalashnikov}}, \bibinfo {author} {\bibfnamefont {E.}~\bibnamefont {Risse}},
  \bibinfo {author} {\bibfnamefont {T.}~\bibnamefont {Sokollik}}, \bibinfo
  {author} {\bibfnamefont {W.}~\bibnamefont {Sandner}}, \bibinfo {author}
  {\bibfnamefont {A.}~\bibnamefont {Andreev}}, \ and\ \bibinfo {author}
  {\bibfnamefont {V.}~\bibnamefont {Tikhonchuk}},\ }\bibfield  {title}
  {\enquote {\bibinfo {title} {Quasimonoenergetic deuteron bursts produced by
  ultraintense laser pulses},}\ }\href {\doibase 10.1103/PhysRevLett.96.145006}
  {\bibfield  {journal} {\bibinfo  {journal} {Phys. Rev. Lett.}\ }\textbf
  {\bibinfo {volume} {96}},\ \bibinfo {pages} {145006} (\bibinfo {year}
  {2006})}\BibitemShut {NoStop}%
\bibitem [{\citenamefont {Kostyukov}\ \emph {et~al.}(2023)\citenamefont
  {Kostyukov}, \citenamefont {Khazanov}, \citenamefont {Shaikin}, \citenamefont
  {Litvak},\ and\ \citenamefont
  {Sergeev}}]{Kostyukov_BullLebedevPhysInst_2023}%
  \BibitemOpen
  \bibfield  {author} {\bibinfo {author} {\bibfnamefont {I.~Y.}\ \bibnamefont
  {Kostyukov}}, \bibinfo {author} {\bibfnamefont {E.~A.}\ \bibnamefont
  {Khazanov}}, \bibinfo {author} {\bibfnamefont {A.~A.}\ \bibnamefont
  {Shaikin}}, \bibinfo {author} {\bibfnamefont {A.~G.}\ \bibnamefont {Litvak}},
  \ and\ \bibinfo {author} {\bibfnamefont {A.~M.}\ \bibnamefont {Sergeev}},\
  }\bibfield  {title} {\enquote {\bibinfo {title} {International exawatt center
  for extreme light studies (xcels): Laser system and experiment program},}\
  }\href {\doibase 10.3103/S1068335623180136} {\bibfield  {journal} {\bibinfo
  {journal} {Bulletin of the Lebedev Physics Institute}\ }\textbf {\bibinfo
  {volume} {50}},\ \bibinfo {pages} {S635--S640} (\bibinfo {year}
  {2023})}\BibitemShut {NoStop}%
\bibitem [{\citenamefont {Danson}\ \emph {et~al.}(2019)\citenamefont {Danson},
  \citenamefont {Haefner}, \citenamefont {Bromage}, \citenamefont {Butcher},
  \citenamefont {Chanteloup}, \citenamefont {Chowdhury}, \citenamefont
  {Galvanauskas}, \citenamefont {Gizzi}, \citenamefont {Hein}, \citenamefont
  {Hillier},\ and\ \citenamefont {et~al.}}]{Danson_HPLSE_2019}%
  \BibitemOpen
  \bibfield  {author} {\bibinfo {author} {\bibfnamefont {C.~N.}\ \bibnamefont
  {Danson}}, \bibinfo {author} {\bibfnamefont {C.}~\bibnamefont {Haefner}},
  \bibinfo {author} {\bibfnamefont {J.}~\bibnamefont {Bromage}}, \bibinfo
  {author} {\bibfnamefont {T.}~\bibnamefont {Butcher}}, \bibinfo {author}
  {\bibfnamefont {J.-C.~F.}\ \bibnamefont {Chanteloup}}, \bibinfo {author}
  {\bibfnamefont {E.~A.}\ \bibnamefont {Chowdhury}}, \bibinfo {author}
  {\bibfnamefont {A.}~\bibnamefont {Galvanauskas}}, \bibinfo {author}
  {\bibfnamefont {L.~A.}\ \bibnamefont {Gizzi}}, \bibinfo {author}
  {\bibfnamefont {J.}~\bibnamefont {Hein}}, \bibinfo {author} {\bibfnamefont
  {D.~I.}\ \bibnamefont {Hillier}}, \ and\ \bibinfo {author} {\bibnamefont
  {et~al.}},\ }\bibfield  {title} {\enquote {\bibinfo {title} {Petawatt and
  exawatt class lasers worldwide},}\ }\href {\doibase 10.1017/hpl.2019.36}
  {\bibfield  {journal} {\bibinfo  {journal} {High Power Laser Science and
  Engineering}\ }\textbf {\bibinfo {volume} {7}},\ \bibinfo {pages} {e54}
  (\bibinfo {year} {2019})}\BibitemShut {NoStop}%
\end{thebibliography}%

\end{document}